\documentclass[manuscript]{aastex62}

\usepackage[normalem]{ulem}

\graphicspath{{./}{figures/}}

\submitjournal{ApJ}

\shorttitle{Multi-slit Approach to Coronal Spectroscopy of MUSE}
\shortauthors{De Pontieu et al.}
\usepackage{natbib}
\begin{document}

\title{The Multi-slit Approach to Coronal Spectroscopy with the Multi-slit Solar Explorer (MUSE)}

\correspondingauthor{Bart De Pontieu}
\email{bdp@lmsal.com}

%\author[0000-0002-8370-952X]{Bart De Pontieu}
\author{Bart De Pontieu}
\affil{Lockheed Martin Solar \& Astrophysics Laboratory,
3251 Hanover St, Palo Alto, CA 94304, USA}
\affil{Rosseland Centre for Solar Physics, University of Oslo,
P.O. Box 1029 Blindern, NO0315 Oslo, Norway}
\affil{Institute of Theoretical Astrophysics, University of Oslo,
P.O. Box 1029 Blindern, NO0315 Oslo, Norway}

%\author[0000-0002-0333-5717]{Juan Mart\'inez-Sykora}
\author{Juan Mart\'inez-Sykora}
\affil{Lockheed Martin Solar \& Astrophysics Laboratory,
3251 Hanover St, Palo Alto, CA 94304, USA}
\affil{Bay Area Environmental Research Institute,
NASA Research Park, Moffett Field, CA 94035, USA}
\affil{Rosseland Centre for Solar Physics, University of Oslo,
P.O. Box 1029 Blindern, NO0315 Oslo, Norway}

%\author[0000-0002-0405-0668]{Paola Testa}
\author{Paola Testa}
\affil{Harvard-Smithsonian Center for Astrophysics,
60 Garden St, Cambridge, MA 02193, USA}

%\author[0000-0002-5608-531X]{Amy R. Winebarger}
\author{Amy R. Winebarger}
\affil{NASA Marshall Space Flight Center,
Huntsville, AL 35812, USA}

\author{Adrian Daw}
\affil{NASA Goddard Space Flight Center,
Greenbelt, MD 20771, USA}

%\author[0000-0003-0975-6659]{Viggo Hansteen}
\author{Viggo Hansteen}
\affil{Lockheed Martin Solar \& Astrophysics Laboratory,
3251 Hanover St, Palo Alto, CA 94304, USA}
\affil{Bay Area Environmental Research Institute,
NASA Research Park, Moffett Field, CA 94035, USA}
\affil{Rosseland Centre for Solar Physics, University of Oslo,
P.O. Box 1029 Blindern, NO0315 Oslo, Norway}
\affil{Institute of Theoretical Astrophysics, University of Oslo,
P.O. Box 1029 Blindern, NO0315 Oslo, Norway}

%\author[0000-0003-2110-9753]{Mark C. M. Cheung}
\author{Mark C. M. Cheung}
\affil{Lockheed Martin Solar \& Astrophysics Laboratory,
3251 Hanover St, Palo Alto, CA 94304, USA}
\affil{Hansen Experimental Physics Laboratory, Stanford University
452 Lomita Mall, Stanford, CA 94305, USA}

%\author[0000-0003-1529-4681]{Patrick Antolin}
\author{Patrick Antolin}
\affil{Department of Mathematics, Physics \& Electrical Engineering, Northumbria University, Newcastle Upon Tyne, NE1 8ST, UK}

\author{the MUSE team}



\begin{abstract}
The Multi-slit Solar Explorer (MUSE) is a proposed mission aimed at understanding the physical mechanisms driving the heating of the solar corona and the eruptions that are at the foundation of space weather. MUSE contains two instruments, a 
multi-slit EUV spectrograph and a context imager.
It will simultaneously obtain EUV spectra (along 37 slits) and context images with the highest resolution in space (0.33-0.4\arcsec) and time (1-4 s) ever achieved for the transition region and corona. The MUSE science investigation will exploit major advances in numerical modeling, and observe at the spatial and temporal scales on which competing models make testable and distinguishable predictions, thereby leading to a breakthrough in our understanding of  coronal heating and the drivers of space weather. By obtaining spectra in 4 bright EUV lines
(\ion{Fe}{9} 171\AA,  \ion{Fe}{15} 284\AA, \ion{Fe}{19}-\ion{Fe}{21} 108\AA) covering a wide range of transition region and coronal temperatures along 37 slits simultaneously, MUSE will be able to ``freeze" the evolution of the dynamic coronal plasma. 
We describe MUSE's multi-slit approach and show that the optimization of the design minimizes the impact of spectral lines from neighboring slits, generally allowing line parameters to be accurately determined.
We also describe a Spectral Disambiguation Code to resolve multi-slit ambiguity in locations where secondary lines are bright. We use simulations of the corona and eruptions to perform validation tests and show that the multi-slit disambiguation approach allows accurate determination of MUSE observables  
in locations where significant multi-slit contamination occurs.
\end{abstract}



\section{Introduction} 
\label{sec:intro}

The physical processes that heat the multi-million degree solar corona, accelerate the solar wind and drive solar activity (CMEs and flares) remain poorly known.   Unfortunately, many of the complex processes in the corona remain invisible to imaging instruments.  Spectroscopic measurements are required, yet the low cadence and small field of view inherent in typical single-slit spectrometers is extremely limiting.    A scientific breakthrough in these areas can only come from radically innovative instrumentation coupled with state-of-the-art numerical modeling.

\subsection{Measurement techniques of plasma conditions in the solar corona} \label{sec:intro_measurement}

The most powerful tool to determine physical plasma conditions through remote sensing is imaging spectroscopy, that is, measuring the spectral radiance over a two-dimensional field of view:
$I(x,y,\lambda,t)$,
where $x$ and $y$ are perpendicular angular or spatial dimensions, $\lambda$ is wavelength, and $t$ is time.
There are a number of ways to accomplish this:
1) Place a position- and photon-energy-sensitive detector at the focal plane of a telescope.
2) Use a telescope with a narrow and tunable spectral passband (or multiple channels with
narrow passbands) to scan spectrally over a field of view (FOV).
3) Use an optical system with spectral dispersion, e.g., a telescope feeding a spectrograph
with a focal plane array.
An important instrumental parameter to consider is the spectral resolving power of the system:
\begin{equation}
R = \lambda / \delta\lambda = \nu / \delta\nu = E / \delta E,
\end{equation}
where $\delta\lambda$, $\delta\nu$, and $\delta E$ are the spectral resolution in terms of wavelength, frequency, and photon energy, respectively. For example, in the extreme ultraviolet (EUV) solar spectrum, which includes the majority of coronal emission, most individual emission lines are clearly resolved with a resolving power of $\sim 2\,000$. As the speed of light is $300\,000$~km~s$^{-1}$, such a resolving power also provides the capability to resolve multiple plasma features along the line-of-sight with velocities that differ by $\sim$150~km~s$^{-1}$ or more, and to centroid velocities with a much greater accuracy. Spatial (angular) resolution is important as well, as averaging the emission from too many individual features can hide the spectral signatures of processes occurring on small spatial scales. For example, clearly blue-shifted spectral profiles, expected during the evaporative phase of solar flares from theoretical models, were not unambiguously identified until the advent of high-resolution observations with the Interface Region Imaging Spectrograph \citep[IRIS,][]{DePontieu:IRIS} in the \ion{Fe}{21} 1354\AA\ line \citep{Young2015}. Dynamic range is a further factor to consider, as the signature of driving mechanisms can have faint signals in close proximity (spectrally and/or spatially) to bright sources.

Given these factors, there are a number of practical trades to consider for the three options listed above. With respect to the solar corona, while the majority of emission is in the extreme ultraviolet
it is worth noting that the emitted radiation can in general span the electromagnetic spectrum from radio frequency to gamma rays. Of the options above, option 1 is perhaps conceptually the most straightforward, as a simultaneous measurement of $I(x,y,\lambda)$ is always achieved. Detectors for hard X-rays (HXR) provide an energy resolution of a fraction of a keV \citep{2019NIMPA.924..321F,2017SPIE10397E..0AA}, 
and a useful resolving power for gamma rays and HXR bremsstrahlung continua, applicable to the very hottest components of the corona, but unfortunately, not for the majority of coronal features that emit EUV. Furthermore, the spatial resolution of state-of-the-art X-ray telescopes \citep{2017SPIE10399E..0JB,2018SPIE10699E..40C} is not sufficient to resolve the cross-sectional profiles  of bright coronal loops, or their footpoints, where the majority of HXR emission is produced.
For soft X-rays (SXR),  \cite{Bandler:2019JATIS...5b1017B} recently demonstrated transition-edge-sensor-based microcalorimeter arrays that provide an energy resolution of 2 eV over the energy range 0.2-7 keV, and thus a resolving power of $3,500$ at 7 keV, but dropping to 100 at $\sim 200$eV ($\sim$ 60\AA).  The count rates for such microcalorimeter systems are limited, and the achievable dynamic range, currently, compromises the utility of such systems for solar observations.

Options 2 and 3 can be implemented in a variety of ways, some of which require a sequence of multiple measurements to obtain a single ``snapshot" of $I(x,y,\lambda)$. Analysis of such data is straightforward when the time scales of the process under study are longer than the cycle time of the sequence. Narrow passband filters to implement option 2 are employed at visible, infrared (IR) and radio-frequency (RF) wavelengths, including the measurement of $I$ for multiple polarization states in order to perform spectropolarimetric inversions using visible and near-infrared (NIR) line emission \citep[e.g.,][]{Scharmer2008}, and microwave and RF observations of gyroresonance emission, which in combination with high resolution EUV observations can provide valuable diagnostics of the 3D structure of the coronal magnetic field \citep[e.g.,][]{2002ApJ...574..453B} and electron beams generated in flares and jets \citep[e.g.][]{2013ApJ...763L..21C}. 
Owing ultimately to the low reflectance and transmittance of materials for EUV wavelengths, the achievable resolving power of EUV multilayer coatings for option 2 is only around 10 to 30.  A spectrograph, option 3, therefore provides the best resolving power in the EUV.

In a traditional imaging spectrograph, 
light passes through a single entrance slit and is dispersed and re-imaged onto the focal plane.  This preserves spatial information in the direction perpendicular to the dispersion, so that one dimension on the detector is spatial and the other spectral. For example, a single exposure provides $I(x,\lambda,t)$ for a single value of $y$. Spectra for a two dimensional field of view can then be obtained by rastering the slit (i.e., re-pointing the telescope) over a range of values of $y$. The overall cadence for a set of exposures from one raster then limits the time scales that can be studied since each $y$ value is obtained at a different time.

A high-cadence alternative to using a raster with a single slit is to use multiple entrance slits to the spectrograph, each of which allows light from the telescope image to pass through for a different value of $y$. Such multi-slit spectrometers have been used at visible and (NIR) wavelengths for solar observations of H$\alpha$ \citep{1974SoPh...37..343M} and \ion{He}{1} 10830\AA\ \citep{2017SoPh..292..158S}.
Independent of details of how the optical system can be implemented for different wavelength regions, there is always the limitation of available detector real estate, that is, the number of resolution elements in the detector. This is the case whether it is a focal plane array or a position-sensitive photon counting system, meaning there is a trade-off between spectral range and spatial coverage. 
At visible and NIR wavelengths, a narrow passband is typically used to limit the spectral range so that there is no overlap of data from adjacent slits on the detector, but this is not in general necessary, just as it is not always necessary to eliminate multiple orders in a traditional single-slit imaging spectrograph \citep[e.g., SUMER.][]{Wilhelm1995}. In fact, the additional spectral information can be quite valuable.

\subsection{Physical conditions in the corona}
\label{sec:intro_reqs}

In order to provide necessary and sufficient observational constraints to determine which physical processes drive solar flares and eruptions, and which are responsible for heating the corona, plasma properties must be measured for multiple temperature regimes at spatio-temporal scales for which competing theories make distinguishable predictions.  Just as the finite number of resolution elements in the focal plane creates a trade-off between spectral and spatial coverage, there is also a trade-off between spectral range, which corresponds to temperature coverage, and spectral resolution. Given the temperature ranges over which the multiple ionization stages of a given element are formed, coronal temperatures may be divided in to three regimes: less than 1 MK - often referred to as transition region (TR) or low coronal temperatures, 1 to a few MK typical of the bulk of coronal structures, and `flare' temperatures approaching 10 MK and beyond. To strike the optimal balance between multi-temperature regime coverage and high spectral resolution, as few as three lines may be chosen to cover these three regimes.

Recent major advances in ``realistic'' numerical modeling can be used to provide direct comparisons with observables, providing a method to directly validate models.
These advances allow a significant improvement over classical approaches in which inversions of spectral data are used to determine ``temperature profiles'' and then comparing those with theoretical predictions. Such an approach often comes with uncertainties inherent to the methodology (e.g., 
Differential Emission Measure (DEM) inversions and non-uniqueness issues,  \citealt{csr122}). A better method lies instead in calculating synthetic observables from advanced forward models of various heating mechanisms and then comparing those with observations of intensity, Doppler shift and non-thermal broadening for a few well chosen spectral lines supplemented with intensities from context images. Such comparisons between synthetic observables and observations, based on a handful of lines, nevertheless allows rigorous tests and improvements of models and provide key insights into the dominant physical mechanisms, e.g., as shown previously with the successful experience from the IRIS satellite \citep{csr79,csr113,Testa:2016ApJ...827...99T,csr123}. 

While the dissipation of magnetic and mechanical energy that drives coronal heating and solar activity likely occurs on plasma scales that cannot be resolved through remote sensing, competing theories do make distinguishable predictions for spatio-temporal correlations between spectral diagnostics (e.g. Doppler speed vs. temperature) at sub-arcsecond spatial scales. In addition, recent observations indicate that loops, the building blocks of the corona, show collective behavior that appears to be mostly resolved on scales of ~400 km \citep{csr4,csr5}. The dynamics of these finely-structured loops have been glimpsed only recently in a 5-minute long time series of images from the High-resolution Coronal Imager (Hi-C) rocket which revealed tantalizing views of braiding \citep{csr6}, but lack the duration, thermal coverage and spectroscopic information necessary to measure line-of-sight (LOS) velocities, and identify non-thermal processes or heating mechanisms. Similarly, measurements of flows and turbulence at high cadence over a wide field of view are key to detect reconnection, waves and plasma flows. Such measurements are critical to determine the initiation mechanisms of flares and CMEs, the role of reconnection in their spatio-temporal evolution, or their interaction with the surrounding corona.

From the dynamic nature of the corona and the mix of small scale heating events with high velocity wave propagation, large flow speeds, and turbulent processes, it is clear that a multi-slit approach will provide a revolutionary view of the physical processes involved. This requires a spectrometer with sufficient spatial and temporal resolution to resolve the cross-sectional profiles and dynamic evolution of velocity and non-thermal motions in bright coronal loops at a broad range of temperatures covering a large area.

The small scale heating processes, whether driven by waves or reconnection, and the need for measuring flows and turbulence during eruptions and flares, underscore the need for multi-slit spectroscopy and imaging, at high cadence (1-20s) and resolution (400 km), covering temperatures from the TR to the hottest parts of the corona. Considering the energy flux necessary to heat the corona, a centroiding uncertainty of 5~km~s$^{-1}$ in the TR and cooler coronal lines and 30~km~s$^{-1}$ for the hotter, flaring, lines is desired. Likewise, a determination of the non-thermal line widths to better than 10~km~s$^{-1}$ in the TR and cooler coronal lines would realize the goal of separating various heating scenarios, while 30~km~s$^{-1}$ will constrain turbulence and waves in flares and eruptions.

The bulk of the plasma in the TR and corona radiates in the EUV providing a sample of 
strong, isolated lines, a significant fraction of which are from various ionization stages of iron covering the full range of temperatures realized in the upper transition region and corona.
Many coronal ions emit at longer wavelengths as a result of forbidden transitions. In contrast to the electric-dipole allowed resonance transitions in the EUV, the forbidden transitions are not strong and isolated relative to cooler lines. In addition, they are not useful for on disk observations when the lines are observed weakly in absorption above strong continuum emission from lower in the atmosphere.

With these scientific goals, constraints and trade-spaces in mind, we have designed the Multi-slit  Solar  Explorer  (MUSE), a high resolution multi-slit spectrometer combined with a context imager.  In this paper, we review the instrument design and discuss analysis techniques specific to multi-slit spectrometers.  We detail the instruments in \S~\ref{sec:muse_intro}.
A key aspect of the multi-slit approach is avoiding multi-slit ambiguity. This aspect of the MUSE design is presented in \S~\ref{sec:purity}. We further investigate the impact of spectral contamination from neighboring slits on accurately measuring spectral line parameters in \S~\ref{sec:impurities}. In \S~\ref{sec:sdc} we provide an overview of the Spectral Disambiguation Code (SDC) developed for the MUSE project, and the various validation tests of this code. We finish the paper with a description of the preliminary development of a deep neural network algorithm to perform multi-slit disambiguation (\S~\ref{sec:sdc_dnn}), and a brief discussion (\S~\ref{sec:discussion}).

\section{MUSE mission}\label{sec:muse_intro}

The Multi-slit Solar Explorer (MUSE) is a proposed space mission that is designed to deliver the high spatial resolution and temporal cadence necessary to understand the basic physical mechanisms that heat the outer solar atmosphere to high temperatures and that drive the eruptions at the foundation of space weather. As with Hinode/EIS \citep{Culhane07:EIS}, this is done by concentrating on highly ionized elements that emit in the EUV, primarily Fe~lines. The MUSE design provides a dramatic advance in very high resolution EUV imaging and spectroscopy of the solar corona at a cadence that is two orders of magnitude higher than previous, current, or pending missions. This is accomplished by using an EUV spectrograph with an innovative 37-slit design and high-throughput spectroscopy in three EUV passbands (108, 171 and 284\AA). 

In order to maximize performance and minimize size, the 108 passband is observed in 2nd order, and the other two are in 1st order. The multi-slit design allows the simultaneous observation of an 170\arcsec $\times$ 170\arcsec\ field of view at a cadence of $\sim~1$ s for an active region (AR).  The slit width and spatial resolution along the slit is 0.4\arcsec, while the slits are separated by 4.45\arcsec. Rastering will allow MUSE to observe the entire active region at high resolution at a cadence of $\sim12$ s (and bright points in quiet Sun and coronal holes at $\sim15-20$~s). In addition to the multi-slit spectrograph, MUSE includes an EUV context imager providing 0.33\arcsec\ resolution images in two passbands, one dominated by the \ion{He}{2} 304\AA\ line formed in the transition region and the other by the \ion{Fe}{12} 195\AA\ line formed in the $\sim1-2$~MK 
corona. These high resolution context images will cover a field of view of 580\arcsec $\times$ 290\arcsec\ at $\sim4$~s cadence for a single passband, and $\sim8$~s cadence for both passbands.

Since both spectral and spatial information will be detected along the dispersion direction, special care must be taken to ensure that disambiguation between the information coming from separate slits is possible. This is achieved by picking strong isolated Fe\ 
lines typical of disparate temperature regions of the solar corona and by tuning the inter slit distance. The three spectral passbands chosen are dominated by spectral lines with wavelengths around 108\AA\ (\ion{Fe}{19} and \ion{Fe}{21}; formed at $\log (T[K]) \approx [7.0,7.1]$), 
around 171\AA\ (\ion{Fe}{9}; $\log (T[K]) \approx 5.9$), and around 284\AA\ (\ion{Fe}{14}; $\log(T[{\rm K}]) \approx 6.4$). The passbands are spectrally wider (respectively, 2\AA, 4\AA, 12\AA) than the (wavelength) separation between neighboring slits, and the multi-slit design can, in principle, lead to overlap of
spectral information from neighboring slits. This is minimized by the selection of narrow passbands to study bright, well-isolated lines as primary diagnostics and the selection of a slit spacing that minimizes possible blends from other slits (see \S\ref{sec:purity}). The present design of MUSE has  a slit spacing of $0.390$\AA\ for the 108 and 171\AA\ bands and $0.780$\AA\ for the 284\AA\ band.  This typically limits multi-slit confusion to regions in which the primary lines are
not bright, or where the plasma has unusual emission measure (EM) distributions (e.g., a predominance of very cool plasma).

\begin{deluxetable}{cccccc}
\caption{Estimated MUSE count rates}
\tablehead{
  \colhead{Target} &  \colhead{\ion{He}{2}~304\AA} & \colhead{\ion{Fe}{9} 171\AA} &
 \colhead{\ion{Fe}{12}~195\AA} & \colhead{\ion{Fe}{15}~284\AA} & \colhead{ \ion{Fe}{19}~108\AA} \\
 & (CI) & (SG) & (CI) & (SG) & (SG) %\\
}
\startdata
AR loops/moss & 125  & 250   & 200    & 200   & -- \\
AR core             & 125  & 50     & 150    & 600   & 25 \\
M2 flare            & 1500 & 7500 & 1.7e5 & 2.5e4 & 3.5e4 / 3400\tablenotemark{a} \\
Microflare         &  900  & 1000 & 1200  & 7000 & 300 / 10\tablenotemark{a} \\
Quiet Sun          & 35     & 35    & 15      & 5       & -- \\
QS Bright Point  & 70     & 50    & 20      & 5       & -- \\
Coronal Hole     & 15     & 10    & 5        & --     & -- \\
CH Bright Point  & 70     & 40    & 15      & 1       & -- \\
\enddata
\label{table:cnts}
\tablenotetext{}{Notes: (1) ``(CI)" and ``(SG)" beneath the spectral line refer to ``context imager" and ``spectrograph" respectively, and their count rate units are photons/s/pixel$^2$ and photons/s/line/pixel respectively; (2) ``pixel" in the units of the estimated count rates refers to the spatial pixel.}
\tablenotetext{a}{Estimate for the \ion{Fe}{21}~108.118\AA\ emission.}
\end{deluxetable}

The high throughput of MUSE is based on large effective areas: for the spectrograph, 3.7~cm$^2$ in the 171\AA\ band, 1.8~cm$^2$ in the 284\AA\ band, and 2~cm$^2$ in the 108\AA\ band, and for the imagers, $0.8$~cm$^2$ in the 304\AA\ band and 5~cm$^2$ in the 195\AA\ band. This will allow high S/N spectroscopic data with exposure times as short as 1s (in ARs), of order $\sim1.5$ seconds in QS and CH bright points (171\AA), and of order 10s in QS and CH (171\AA). These estimates are based on count rates listed in Table~\ref{table:cnts} that have been obtained from the analysis of a variety of available coronal data, as detailed in Appendix~\ref{appendix:cnts}. 

Advanced numerical modeling forms an integral part of the MUSE science investigation. As stated in the preceding section, comparisons between observations and synthetic observables from the models will provide unprecedented constraints on theoretical models, allowing us to distinguish between competing models and determine which models work, and which do not. Numerical models also enable confident interpretation of the complex observables provided by the instrument and indeed emitted by the physical processes ruling coronal physics.  

\begin{figure*}
    \centering
    \includegraphics[width=0.8\textwidth]{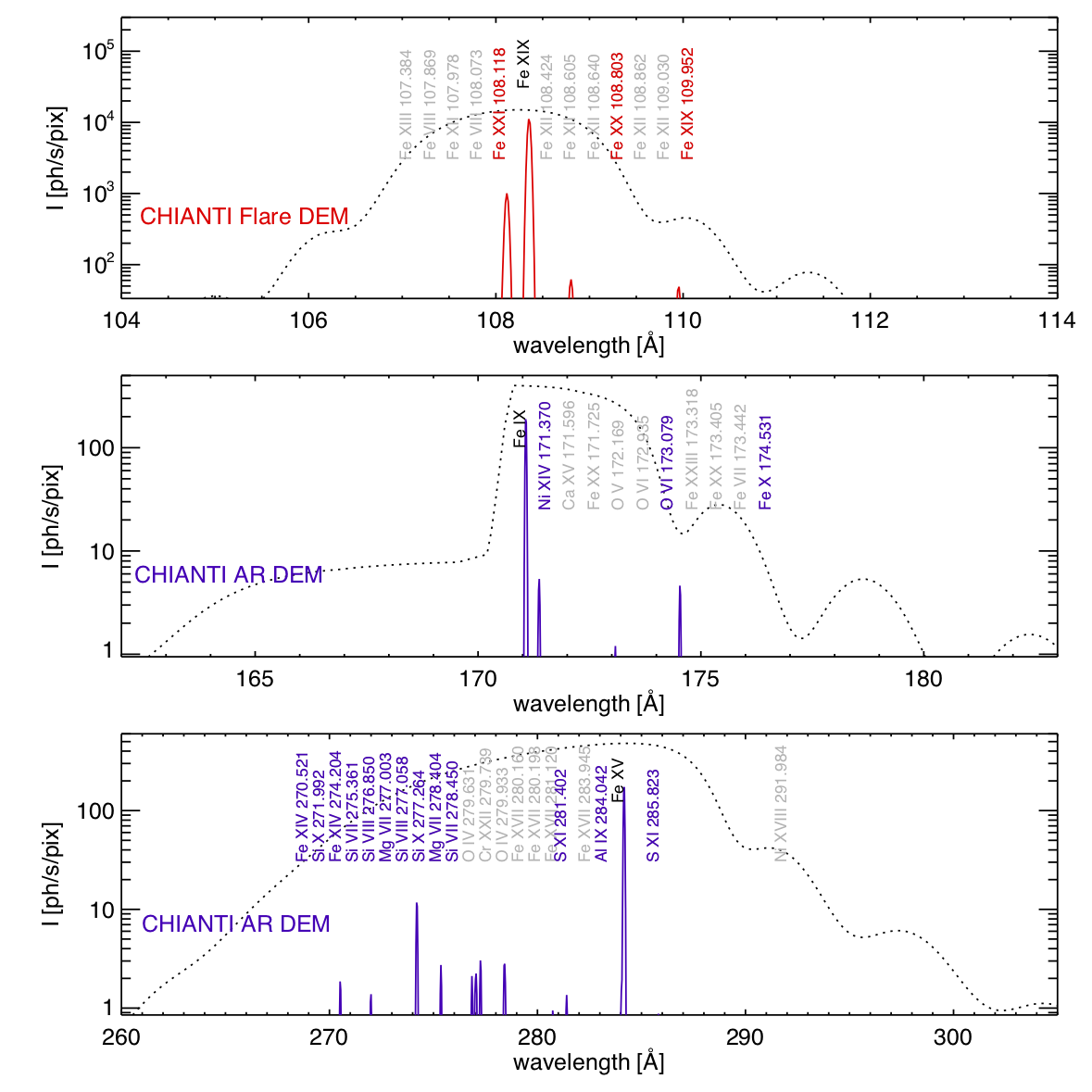}
    \caption{The MUSE spectral bands are centered around bright, isolated spectral lines, as shown in these MUSE synthetic spectra (colored solid lines) created by using the CHIANTI flare DEM (top panel, for the 108\AA\ passband) and active region DEM (middle and bottom panels, for the 171\AA\ and 284\AA\ passbands, respectively), convolved with the MUSE effective areas (shown, scaled, as dotted lines). Lines with intensity within 3 orders of magnitude of the main line are marked by colored labels, while weaker lines are marked by light gray labels.}
    \label{fig:fig1}
\end{figure*}

\section{Spectral Purity} \label{sec:purity}

One potential issue with the multi-slit spectrometer design is that spectra originating from different slits can overlap in the focal plane, creating confusion 
between the spatial and spectral information.  In this section, we discuss the selection of the MUSE instrumental parameters to minimize this overlap.
The primary tool to evaluate possible overlap starts with the synthesis of MUSE spectra from numerical models, as observations of the temperature and velocity structure of the corona at the MUSE resolution simply do not exist. 

A useful measure of this overlap is the spectral purity of MUSE lines, that is, the fraction of detected light within an observed MUSE line profile that is from those wavelengths (rather than from neighboring slits and thus other wavelengths).  We conclude this section 
using a series of numerical models to predict the spectral purity of MUSE data for a variety of solar scenes. The quantitative effect of spectral impurities on measured line parameters is evaluated in \S~\ref{sec:impurities}.

\subsection{Selection of Passbands for the Main Lines}\label{se:purity_chianti}

To minimize the impact of overlapping spectra, the MUSE spectral passbands have been chosen to include bright EUV lines that are spectrally isolated as much as possible. This is illustrated 
by the MUSE synthetic spectra in Figure~\ref{fig:fig1}, which shows 
in a logarithmic plot that the dominant lines are significantly brighter than any secondary lines by an order of magnitude or more, for typical conditions on the Sun: AR --- for the 171\AA\ and 284\AA\ passband --- and Flare --- for the 108\AA\ passband --- DEMs from CHIANTI \citep{Dere:1997,Landi2012,DelZanna:2015}.

MUSE will be able to measure the properties (intensity, velocity, width) of the dominant lines (\ion{Fe}{9} 171\AA, \ion{Fe}{15} 284\AA\ and \ion{Fe}{19} 108\AA) within $\sim 1.8$~s where these lines are bright. As described in the rest of the paper, this can be readily accomplished without significant multi-slit ambiguity under typical conditions. 

\begin{figure*}
    \centering
    \includegraphics[width=0.8\textwidth]{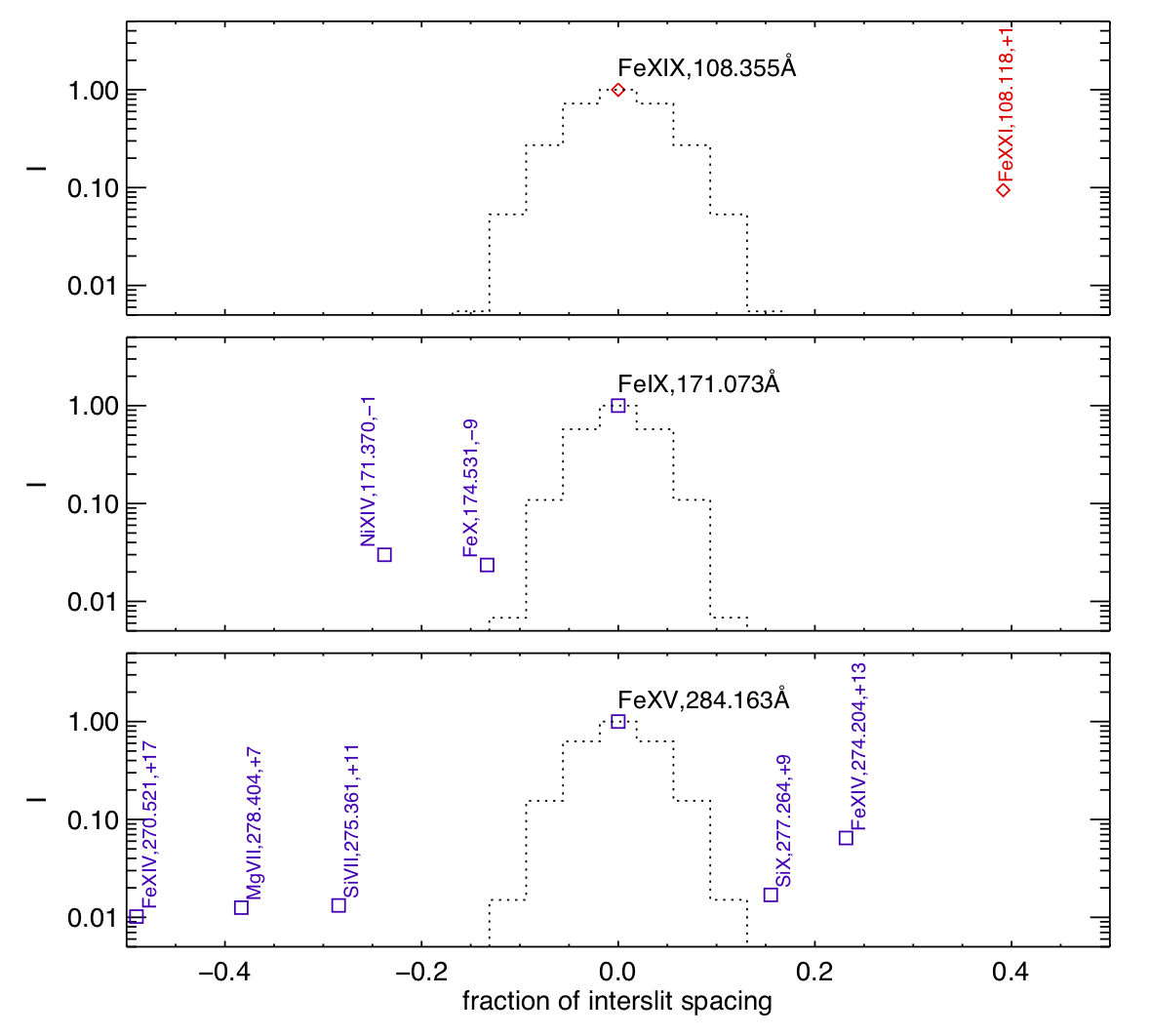}
    \caption{The inter-slit spacing in MUSE's multi-slit design has been chosen to minimize contamination of the main line by spectral lines from neighboring slits. The x-axis of these plots corresponds to the wavelength relative to the main line rest wavelength, normalized to the inter-slit spacing. The y-axis shows the relative strength of spectral lines to the main line, assuming the same DEMs (and convolution with MUSE effective areas) as in Figure~\ref{fig:fig1}. Secondary lines are labeled with the slit number (relative to the slit of the main line) from with they originate, e.g., the \ion{Fe}{14} 274.204+13 label on the bottom panel indicates the location where the \ion{Fe}{14}~274\AA\ line from 13 slits to the right falls. The dotted lines show the thermally and instrumentally broadened line profiles for the main lines.}
    \label{fig:fig2}
\end{figure*}

\begin{figure*}
    \centering
    \includegraphics[width=0.8\textwidth]{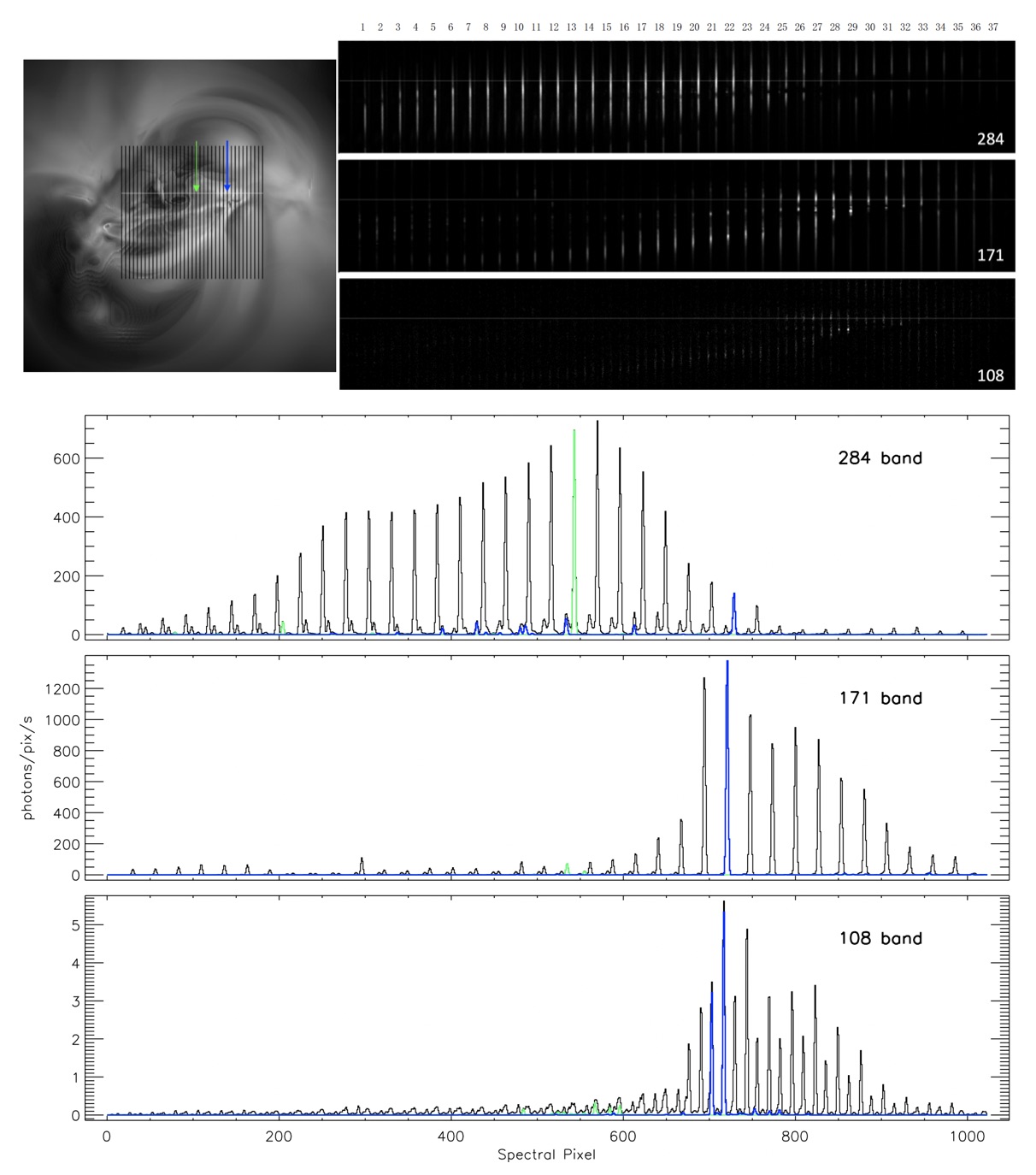}
    \caption{MUSE data products of model A, a quiescent AR simulation \citep{mok2008}, illustrate that under typical conditions and in most locations, there is no significant overlap of spectra. The image in the upper left shows a synthetic \ion{Fe}{9} 171\AA\ image with slits overlaid. The three panels in the upper right show the MUSE observables in all three passbands. The bottom three rows (284\AA, 171\AA, and 108\AA) show MUSE synthetic spectra from a 
    horizontal cut (white line in top-left panel) 
    through the active region. The 284 and 171\AA\ channels are dominated by \ion{Fe}{15} and \ion{Fe}{9}, respectively. Black lines show the total spectrum, while green and blue lines show the contributions from two individual slits. The simulated active region does not reach high enough temperatures to show significant \ion{Fe}{19} 108\AA\ emission. Instead the emission shown in the 108\AA\ channel is dominated by \ion{Fe}{8} (see Figure~\ref{fig:fig1}), which however shows very low counts (of order a few counts/s). 
    }
    \label{fig:fig3}
\end{figure*}

\subsection{Inter-slit Spacing}\label{sec:purity_inter-slit}
The dominant bright lines (\ion{Fe}{9} 171\AA, \ion{Fe}{15} 284\AA, and \ion{Fe}{19} 108\AA), 
passbands and inter-slit spacing have been chosen to reduce the effects
of overlapping slit spectra to an absolute minimum. Even though,
as shown in Figure~\ref{fig:fig1}, for typical conditions the main lines are predicted to be much brighter than secondary lines, it is of course possible that some of these weak lines from neighboring slits appear in the vicinity (on the detector) of the main lines.
Therefore, in order to minimize the effects of overlapping slit spectra we have explored in detail a large range of inter-slit spacing values by synthesizing MUSE spectra from a wide sample of DEMs and synthetic profiles from numerical simulations (\S~\ref{sec:purity_sims}), and then analyzing the amount of contamination of the main lines.

As a result of this analysis, the inter-slit spacing has been carefully tuned (0.390\AA\ in second order for 171 and 108\AA, 0.780\AA\ in first order for 284\AA) to ensure that the weak lines will, under most conditions, not blend with the main lines. This is illustrated in Figure~\ref{fig:fig2}, which shows where on the detector the secondary lines are expected to fall, from which slit they originate, and what their relative intensity is compared to the main line.

\begin{figure*}
    \centering
    \includegraphics[width=0.8\textwidth]{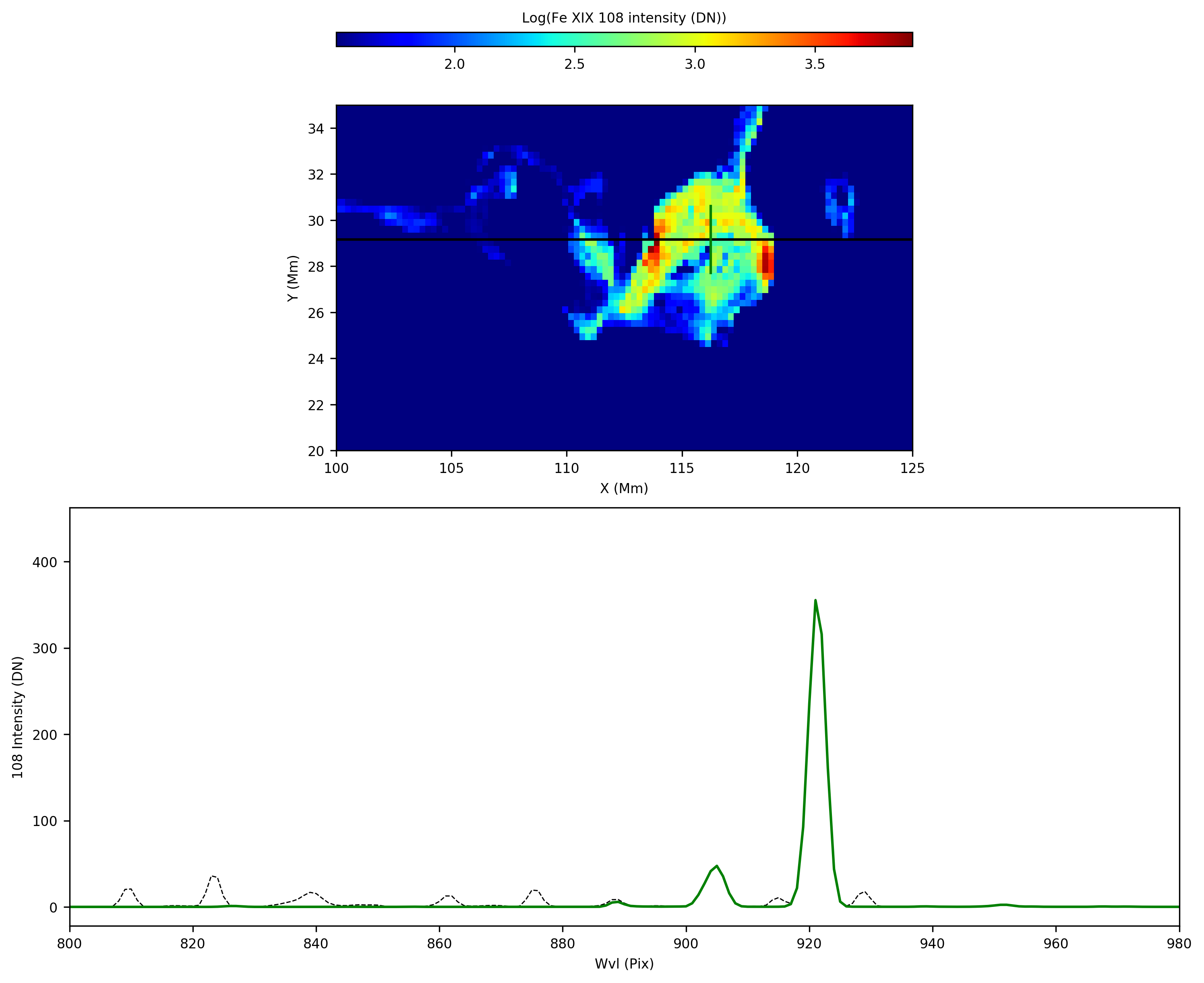}
    \caption{Synthetic MUSE spectra from a horizontal cut (bottom panel) through model C, a MURaM simulation of a C-class flare \citep{CheungRempel:2018}.  Top panel shows \ion{Fe}{19} 108\AA\ synthetic MUSE image. The bottom panel shows MUSE spectrum in black dotted lines, while green full line shows the contribution from one individual slit which shows, from right to left, \ion{Fe}{19}, \ion{Fe}{21} and \ion{Fe}{8}  (see Figures~\ref{fig:fig1} and~\ref{fig:fig2}), cleanly separated from spectra of neighboring slits. The numerical domain of this simulation is smaller than the full MUSE FOV. We tiled the simulation in a periodic fashion to cover the full FOV. Note that this represents a worst case scenario in terms of potential multi-slit ambiguities, since it implies that MUSE would be observing two flares occurring at the same time within its FOV.}
    \label{fig:fig6}
\end{figure*}

\subsection{Spectral Purity Calculations for 3D Radiative MHD Models}\label{sec:purity_sims}

It is useful to define a quantitative measure with respect to the possible overlap of line profiles from different slits:  spectral purity - the fraction of light within the  $\pm 2$ pixels of a line that is from the associated slit, or in other words, the fraction that is not from neighboring slits. While each primary line for MUSE is by far the dominant line within its respective passband (Fig.~\ref{fig:fig2}), intensity and velocity gradients between locations and temperature regimes must be considered in order to quantify and mitigate possible effects of overlapping spectra from adjacent slits.  To study the effect of this in detail, we have used advanced numerical simulations of quiet Sun, quiescent active regions, emerging AR, and flares.  
To synthesize MUSE spectra, all emission lines from all slits are folded through the effective area of each channel, convolved with the instrumental resolution, and the total signal on the detector from all 37 slits is calculated (using the CHIANTI database). Further details may be found in appendix~\ref{sec:sdc_response} and an example is shown in Fig.~\ref{fig:fig3}.

The simulations selected for this study are focused on reproducing the typical conditions of the solar atmosphere. They have been extensively
compared and tested for this purpose 
\citep[e.g.][]{Olluri:2015fk,Testa:2016ApJ...827...99T,Winebarger:2016ApJ...831..172W,CheungRempel:2018,Carlsson-Public-Bifrost2016,Hansteen2019, Antolin2017}.
Throughout the rest of the paper, we will use three of these numerical simulations to illustrate various issues related to multi-slit effects:
\begin{enumerate}
    \item Model A: a 3D hydrodynamic simulation from Predictive Science Inc of a quiescent active region based on the observed magnetic field of NOAA Active Region (AR) 7986 \citep{mok2005,mok2008}. The properties of this simulation have been found to agree well with observations \citep{mok2016,Winebarger:2016ApJ...831..172W}.
    \item Model B: a magnetic flux emergence simulation with fairly strong ambient field using the Bifrost code \citep{Hansteen2019}.  Multiple reconnection events occur as the field breaks through the photosphere and expands into the outer atmosphere. This reproduces many complex chromospheric and transition region observables associated with UV bursts, moss and flux emergence. The treatment of the lower atmosphere includes most of the relevant physical processes at the expense of a rather small numerical domain. Consequently, in order to reproduce a MUSE FOV, this model has been tiled seven times. This strong emerging region shows strong low atmospheric activity (from reconnection) as well as bright upper TR moss, both leading to bright TR lines. These TR lines are the most common secondary lines in the MUSE passband. By tiling this simulation over the spatial extent of a whole AR, this case thus represents an absolute worst case scenario in terms of multi-slit contamination from TR spectral lines.
    \item Model C: the first 3D radiative MHD simulation of a flare inspired by the observed evolution of NOAA AR 12017, using the MURaM code \citep{CheungRempel:2018}. A new flux tube emerges in the vicinity of a pre-existing sunspot producing several solar flares with energies equivalent to C-Class flares. This model is, for the first time, able to reproduce many high energy observables. Similar to the previous case, the FOV of this model has been tiled two times in order to match MUSE FOV. Note that tiling the numerical domains represents a worst case scenario in terms of potential multi-slit ambiguities, since it implies that MUSE would be observing two flares at the same time within its FOV.
\end{enumerate}  

\begin{figure*}
    \centering
    \includegraphics[width=0.9\textwidth]{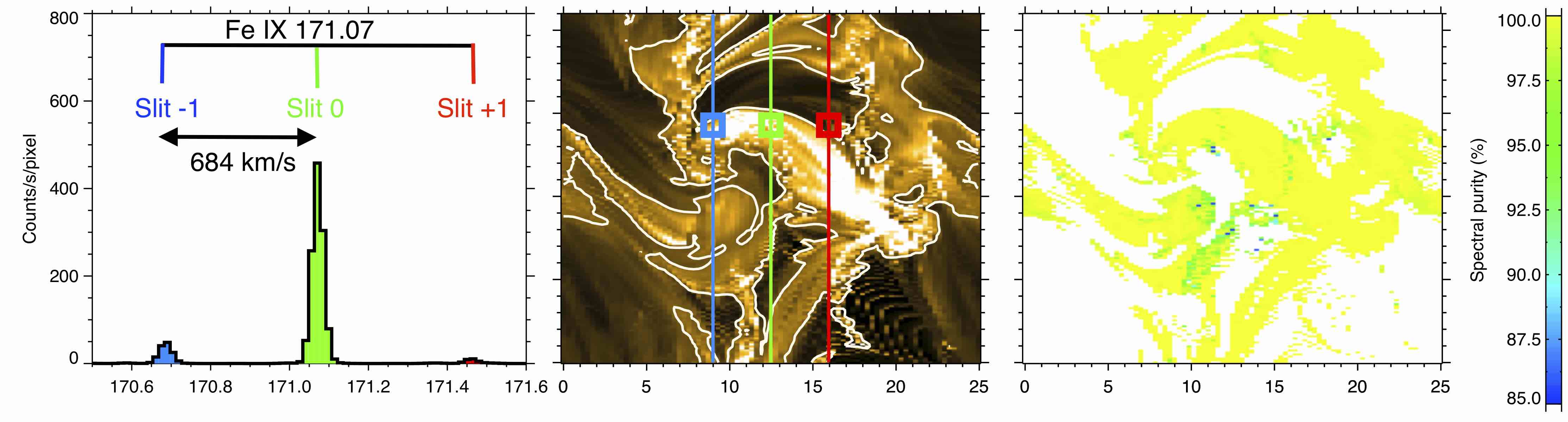}
    \includegraphics[width=0.9\textwidth]{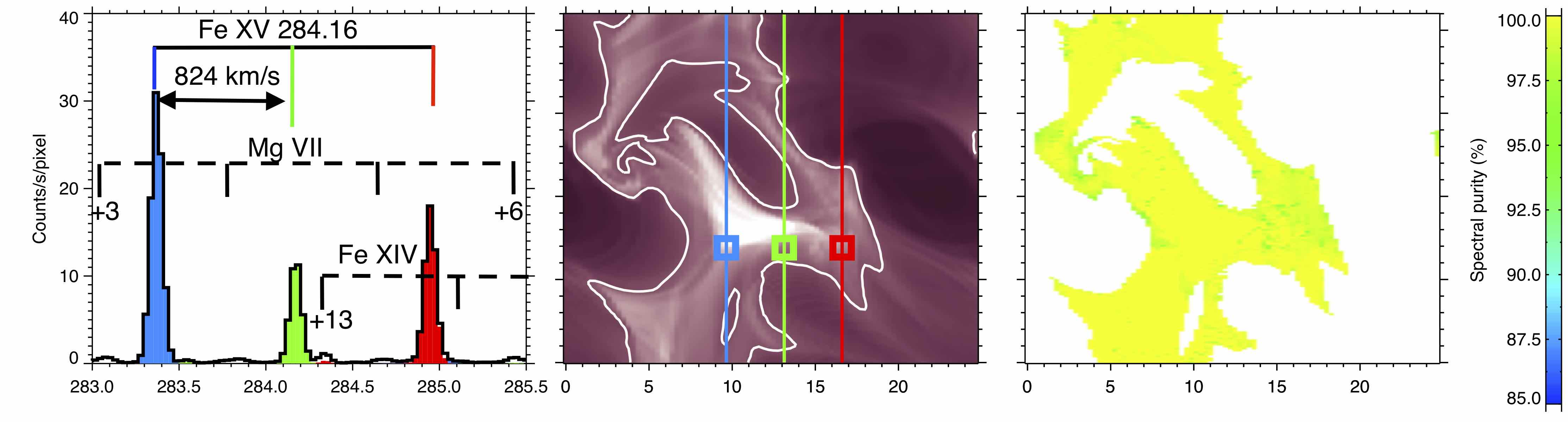}
    \includegraphics[width=0.9\textwidth]{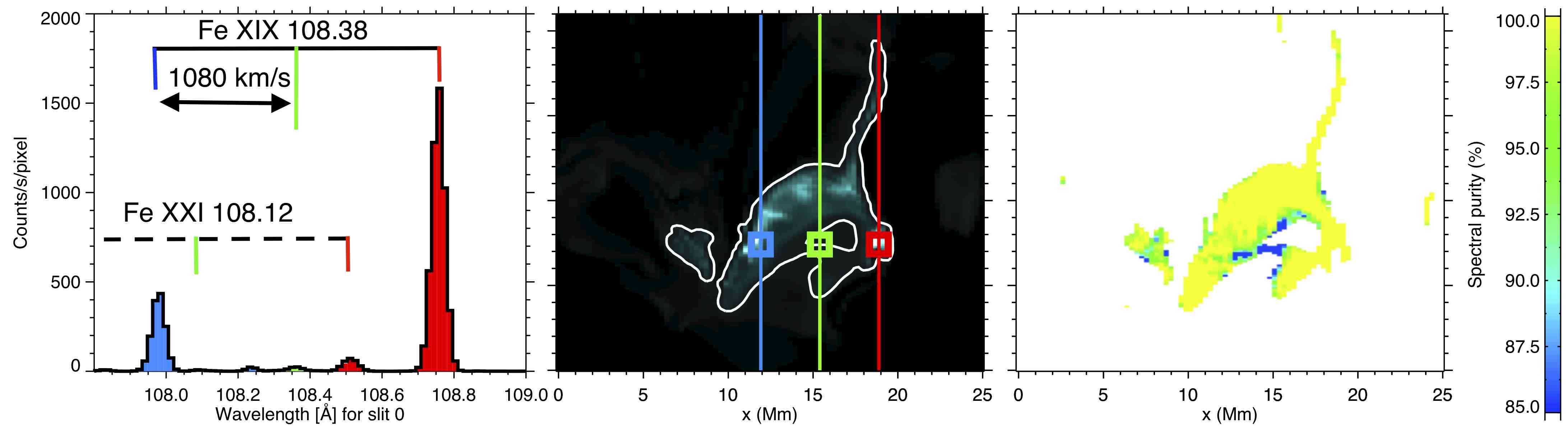}
    \caption{The MUSE spectral lines are well isolated for most locations in typical MUSE targets (with $S/N>10$, white contours in middle column) as evidenced by maps of spectral purity (right column) for \ion{Fe}{9}~171\AA\ (top) and for \ion{Fe}{15}~284\AA\ (middle) using the flux emergence simulation of model B \citep{Hansteen2019} and \ion{Fe}{19}~108\AA\ (bottom) using the flare simulation of model C \citep{CheungRempel:2018}. Examples of a typical spectrum (left column) show well-isolated main lines and minor contamination from secondary lines. The slit number from which the spectral contribution originates is indicated, e.g, ``+13" \ion{Fe}{14} arises from a slit 13 slits to the right of the central slit 0. The spectral purity (fraction of 
    intensity within  $\pm 2$ pixels of MUSE detected line that is not from neighboring slits) is close to 100 in most locations.
    \label{fig:fig9_newversion}}
\end{figure*}

To illustrate how clean the MUSE spectra are, we show several line plots from a simulated MUSE data product (which combines spatial and spectral information). Figure~\ref{fig:fig3} shows spectral line plots for a horizontal cut through model A. In most locations in this region, there is no significant overlap of main lines and secondary lines from neighboring slits. Where there is some overlap, the contaminant is usually a minor contribution to the main line, or the main line and contaminant are both weak. A similar picture emerges for a cut through model B (Fig. \ref{fig:fig9_newversion}, top two rows), and the flare of model C (Fig.~\ref{fig:fig6} and bottom row in Fig.~\ref{fig:fig9_newversion}). More examples can be found in appendix~\ref{sec:ap_prof}.  Models A and B do not reach high enough temperatures to produce any significant \ion{Fe}{19}. Consequently, the 108 channel shows mostly very faint \ion{Fe}{8} emission (which is nevertheless well isolated). We note that in the hot core of non-quiescent ARs, MUSE will be able to detect \ion{Fe}{19} 108\AA\ emission (See Table~\ref{table:cnts} and \citet{2014ApJ...790..112B}\footnote{The EUNIS sounding rocket clearly observed \ion{Fe}{19} 592~\AA\ throughout AR11726, with a similar effective area and resolving power, but an order of magnitude coarser spatial resolution than MUSE.}).

We have also performed spectral purity calculations. These demonstrate the paucity of multi-slit contamination or ambiguity in MUSE data, which is the result of a careful selection of strong, isolated coronal emission lines to target, the passbands to observe those lines, and the inter-slit spacing.
Figure~\ref{fig:fig9_newversion} illustrates this (right column) for the flux emergence simulation of model B (top two rows) and the flare simulation of model C (bottom row). Both represent, in some sense, worst case scenarios, as described above. Nevertheless the spectral purity is typically close to 100\%, i.e., very little contamination (which is defined as 100\%-spectral purity). In some locations, contamination of order a few \% can occur. The left panels reveal how the spectra are cleanly isolated for 3 neighboring slit positions.

Spectral purity provides a valuable metric with which to evaluate and optimize instrumental parameters in order to limit and largely avoid spectral overlap of line profiles with those from neighboring slits.  A critical evaluation of the effect on measured line-of-sight motions for the main lines due to any residual ``spectral impurity" is given in \S~\ref{sec:impurities} below.

\section{Impact of Spectral Impurities}
\label{sec:impurities}

In this section, we consider the impact of the small residual contamination due to overlapping spectral windows.
To determine to what extent such contamination would affect our ability to centroid and determine the line width accurately for the main lines in the MUSE passbands, we performed Monte Carlo simulations that take into account the characteristics of the MUSE instrument (including photon and readout noise). In particular, we investigate the signal-to-noise that is required in the MUSE spectral observations to determine the Doppler shift and line width within the desired uncertainty, both in the case of no significant contamination and in the presence of a contaminant.
We simulated MUSE \ion{Fe}{9}~171\AA , \ion{Fe}{15}~284\AA , and \ion{Fe}{19}~108\AA\ spectra for different signal-to-noise levels (from 10 to 1600 photons, as total line intensity, which corresponds to S/N=3 to 40), running 1000 Monte Carlo simulations for each case, and then estimated the uncertainty in the determination of line shift and width from these 1000 random realizations.
We included photon (Poisson) noise by generating photon counts with the IDL Poisson number generator, added gaussian readout noise assuming the worst-case value of 20 $e^{-}$ RMS,
and modeled the spectral lines as Gaussian profiles including instrumental broadening, and thermal broadening (the 1/e thermal width $w_{\rm th}$ is $\sim$ 16, 27, and 52~km~s$^{-1}$, for \ion{Fe}{9} 171\AA, \ion{Fe}{15} 284\AA, and \ion{Fe}{19} 108\AA, respectively). We also added a non-thermal broadening that we assumed to be 15~km~s$^{-1}$ (a typical value for coronal conditions, see e.g., \citealt{Brooks16,Testa:2016ApJ...827...99T}). We fit the simulated profiles including noise with a Gaussian line profile (using the {\tt mpfitpeak} IDL routine). 

\begin{figure*}
    \centering
    \includegraphics[width=0.8\textwidth]{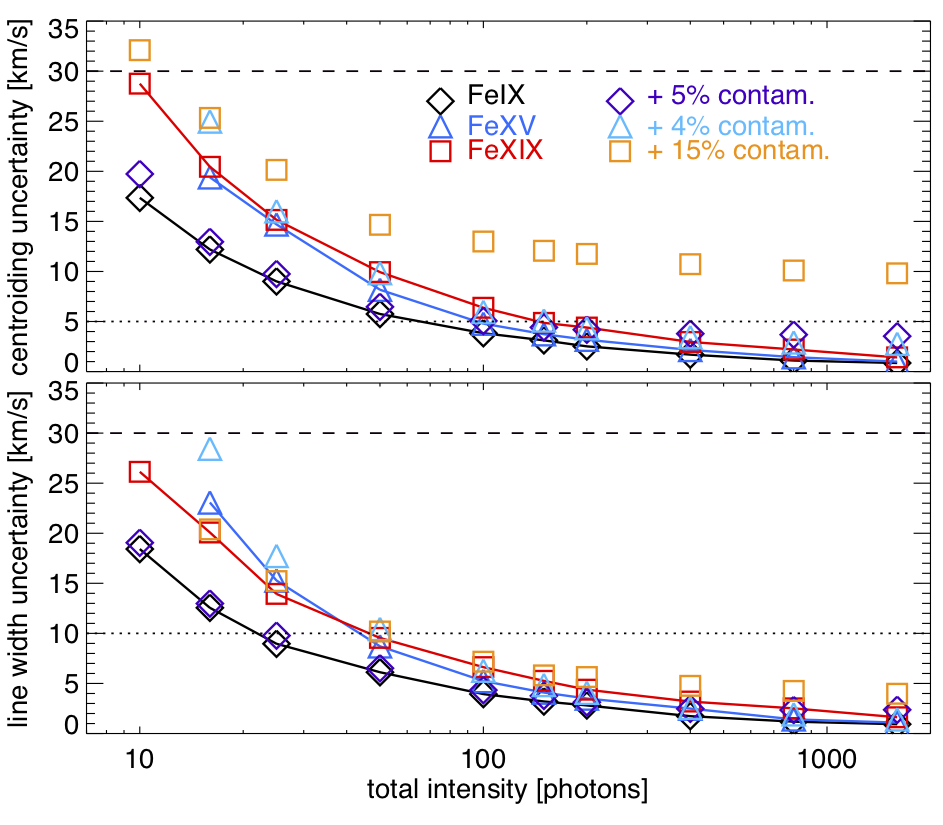}
    \caption{Estimate of uncertainties in centroiding (top panel) and line width (bottom panel) determination in MUSE, from Monte Carlo simulations, for \ion{Fe}{9}~171\AA\ (black diamonds, solid black line), \ion{Fe}{15}~284\AA\ (blue triangles, solid blue line) and \ion{Fe}{19}~108\AA\ (red squares, solid red line), as a function of total intensity of the line (and therefore signal-to-noise). The lighter colored 
     (purple, light blue, orange, for 171\AA, 284\AA, and 108\AA\ respectively) symbols represent the corresponding estimates when a contaminant (with intensity relative to the intensity of the main line, as indicated in the inset) is present (we show the maximum value for our sampled velocities; see text for details). Note that the level of contamination (5\%, 4\%, 15\%) quoted is based on the same definition as that of the spectral purity of Fig.~\ref{fig:fig9_newversion} (i.e., measured within $\pm$ 2 pixels of the main line center, and spectral purity equal to 100-contamination). 
   The dotted lines show the maximum desired uncertainty for the 171\AA\ and 284\AA\ lines (5 and 10~km~$s^{-1}$ for Doppler velocity and
    line width respectively), and the dashed lines for the 108\AA\ line (30~km~$s^{-1}$ for both line shift and width).} 
    \label{fig:fig10}
\end{figure*}

The results are summarized in Figure~\ref{fig:fig10} (symbols connected by lines, for the case without multi-slit contamination), which shows the estimated level of uncertainties (darker color symbols with connecting solid lines) on line shift and (1/e) width, for the three main lines, as a function of total intensity in the main lines (and therefore as a function of signal-to-noise). The Monte Carlo simulations 
show that the centroiding and line width can be measured within the desired uncertainty, for S/N $\sim 8$, 10, 3, for 171\AA, 284\AA\ and 108\AA, respectively.

To obtain a rough idea of the effects of multi-slit contamination and/or how often a single Gaussian fitting approach (without considering 
multi-slit contamination) can be used, we performed additional Monte Carlo simulations, in which we investigate the effect of an unresolved contaminant on the measurement of line shift and width. We added a contaminant, which we assumed to be instrumentally broadened, with a thermal width of $w \sim 15$~km~s$^{-1}$ (e.g., an Fe line formed at $\log T[K] =5.9$) and a non-thermal line width of 15~km~s$^{-1}$. We investigated the effect of the contaminant as a function of its relative intensity with respect to the main line, and the relative velocity with respect to the main line, within 
$2$ 
pixels from the center of the main line, since we want to explore the effect of an unresolved line. In the presence of a contaminant with a certain offset with respect to position of the main line, the determination of the main line shift and width will be affected by both a systematic error (bias, caused by the shift in the average value of the sample from the uncontaminated case), and a statistical error (given by the standard deviation, as done above for the case without contaminant). We estimate the maximum uncertainties (in the derivation of line parameters) that take into account both the systematic and statistical effects. In order to do so, we derived (for each set of parameters) the range of values, symmetric with respect to the true values of the line properties, in which the line shift/width falls with a 
$\sim 68\%$ 
probability (one sigma). For this calculation, we take into account the shifted probability distributions, as derived from average and standard deviations of the Monte Carlo simulations. We note that the maximum effect is found for the largest velocity offset within the $\pm~2$ pixel range. It is this maximum uncertainty that we plot below, as a worst case.

We note that if the lines are separated by more than 2 spectral bins, and the contaminant has significant intensity, this secondary line would be evident and/or flagged by the level 2.5 MUSE data product in the data pipeline that uses the SDC to flag locations with significant multi-slit ambiguity (\S~\ref{sec:discussion}). In such a case, a single Gaussian fit would not be performed. Instead a fit with two components would be carried out, or the SDC would be used to disambiguate the data. 

The results from the Monte Carlo simulations including a contaminant are  shown in Figure~\ref{fig:fig10} (symbols without connecting lines). We find that a contaminant with total intensity 10\%, 8\%, and 30\% of the total intensity of the main line, for \ion{Fe}{9} 171\AA\ (at S/N=10), \ion{Fe}{15} 284\AA\ (at S/N=12), and \ion{Fe}{19} 108\AA\ (at S/N=4) respectively, shifted by 2 pixels with respect to the main line (i.e., worst case, since the largest effects are for the largest line shifts), would not significantly impact our ability to determine the main line parameters. At a line shift of 2 pixels, these levels of total contamination correspond to a 
spectral purity (within $\pm 2$ pixels of the main line) of about 95\%, 96\%, and 85\% (for 171\AA, 284\AA, and 108\AA\ respectively), as a line shift of 2 pixels means only $\sim$ half of the contaminating line is within $\pm 2$ of the main line. These values of spectral purity thus correspond to a level of contamination within $\pm 2$ pixels of the main line of 5, 4, and 15\% (and total contamination of 10, 8, and 30\%).  The spectral purity
in Fig.~\ref{fig:fig9_newversion} 
shows that for typical bright MUSE targets the contamination is generally expected to be below these levels of contamination for typical solar conditions. 

\begin{figure*}
    \centering
    \includegraphics[width=0.8\textwidth]{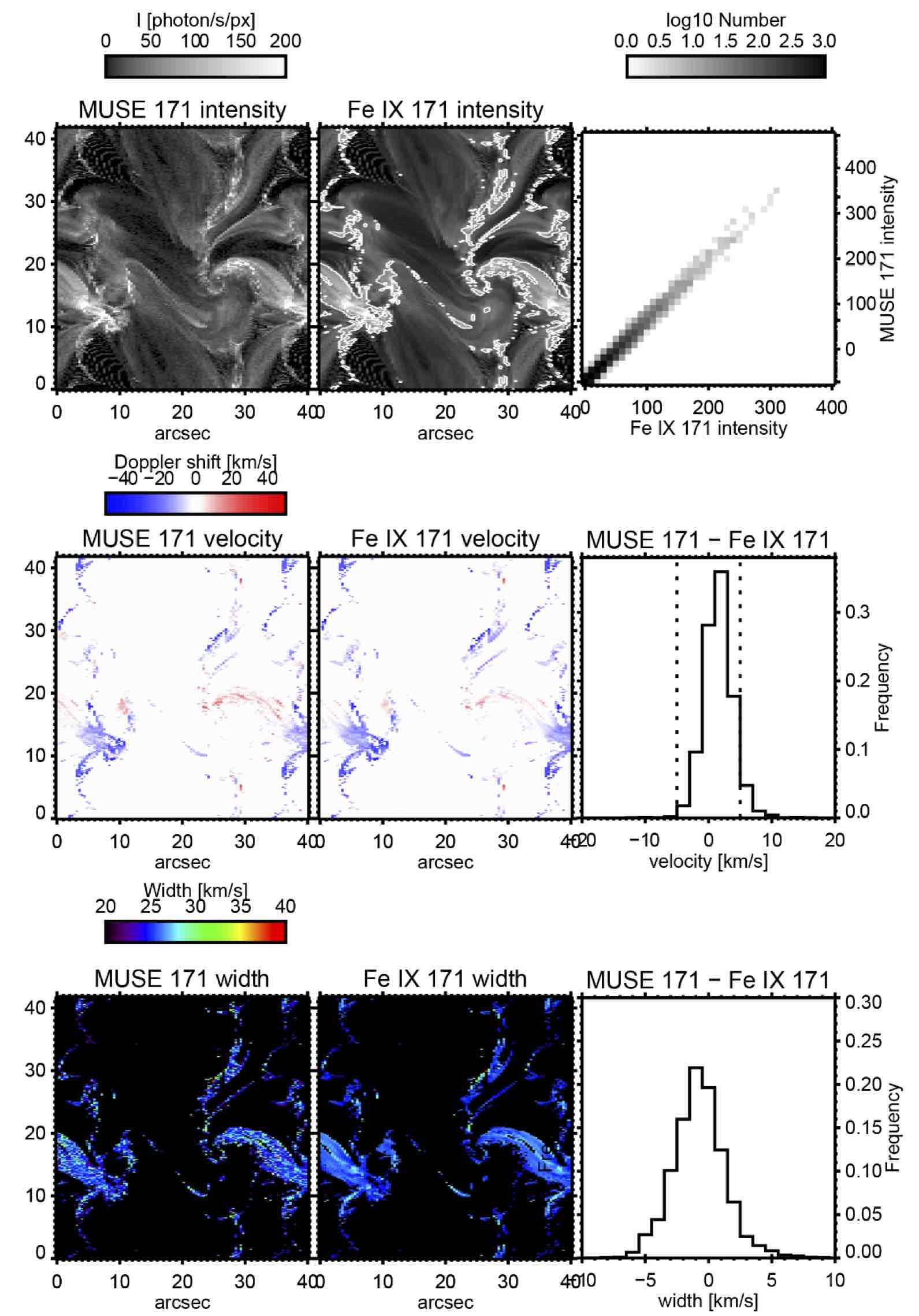}
    \caption{Single Gaussian fits to MUSE observables (left column) from the flux emergence simulation in model B \citep{Hansteen2019} reproduce intensity, velocity 
    and broadening of the ground truth (middle column) to within the desired uncertainty in almost all locations (with ${\rm S/N} > 10$, white contours in intensity).
    }
    \label{fig:fig11}
\end{figure*}

The results of these Monte Carlo simulations are confirmed by the results of  single Gaussian fits to synthetic MUSE spectra from advanced numerical simulations, in this case model B 
(shown also in top and middle row of Fig.~\ref{fig:fig9_newversion}). We calculated synthetic MUSE spectra including photon noise and all significant contaminants. In regions where the MUSE spectral line is bright, we performed single Gaussian fits, and compare the derived line properties to the ground truth from the simulations.  Figure~\ref{fig:fig11} shows that the line shift (second row) and width (third row) are typically determined within the desired uncertainty for most locations. In the following sections we discuss additional strategies to address the effect of contamination for the limited number of locations where the determination of line properties cannot be performed within the desired uncertainty.

All of the above results are thus for a nominal approach in which a single Gaussian fit is used and the Spectral Disambiguation Code (SDC, see \S~\ref{sec:sdc}) is not applied to derive the main line parameters. A key point here is that multi-slit disambiguation is not required for most conditions, and the MUSE data pipeline will flag locations where ambiguity does exist.

\section{Spectral Disambiguation Code}
\label{sec:sdc}
While the spectral purity will be high in the majority of data acquired by MUSE, there are some locations and conditions in which disambiguation can help identify the main lines and isolate the contaminant components. It is for these conditions that we have developed the SDC. The aim of this code is to identify and characterize occurrences in the MUSE data in which multi-slit ambiguities may be present. It will also be provided to the community to help decompose the MUSE spectra.

The SDC solves, as an intermediate step, for the emission measure (as a function of temperature, velocity, and slit position) that reproduces the multi-slit spectrum accurately. This intermediate product is not the end goal of the SDC, but only used to find the best fit to the MUSE multi-slit spectrum. The SDC flags
locations of possible multi-slit confusion and identifies the main and secondary lines (including slit number) to isolate, for each slit position, contributions from secondary lines from neighboring slits.

In Section~\ref{sec:sdc_principles} we describe the 
principles of the SDC algorithm.  We have tested this code in great detail using several different advanced numerical simulations with different solar scenes. In the first series of tests, we determined how well the code alone can return the spectrally pure intensities and velocities of the three 
main spectral lines; we discuss these tests in detail in \S~\ref{sec:sdc_target}.   Our current baseline approach, however, is not to use the SDC to determine the parameters of the main lines, but instead to use the SDC to identify where the main line emission occurs on the detector and isolate any contamination from secondary lines. This method is demonstrated in Section~\ref{sec:sdc_baseline}.  The contaminants can then either: (a) be subtracted from the MUSE data, or (b) the SDC information on the contaminants can be used as an initial value for multiple component fitting.  Our preferred approach is for analysis of the MUSE spectra to occur on the original data of the main line rather than a derived product, but both are possible, and up to the end user.  We have found that the MUSE data can consistently be successfully disambiguated where multi-slit ambiguities exist. Detailed tests show that the SDC inversion performs very well, even in locations where the main lines are weak and secondary lines such as \ion{Fe}{10}~174.53\AA , \ion{Fe}{8}~108.07\AA\ or \ion{Si}{10}~277.26\AA\ become similar in intensity to the main lines.  

\subsection{Principles of SDC} \label{sec:sdc_principles}

The method underlying the SDC is detailed in~\citet{Cheung:Multicomponent}. Here we summarize briefly the principle behind the method. Consider a unit (emission measure) of solar plasma at some temperature 
($\log{T}$) and Doppler velocity ($v$) in the FOV of one of the MUSE slits. 
To compute the resulting MUSE spectrogram on the detector, we first use CHIANTI to compute the emission spectrum, including all spectral lines that could fall on the detector from any of the 37 slits, that is, a wavelength range of (in principle) +/- 36 times the spectral inter-slit spacing, as well as thermal Bremsstrahlung. We then fold the spectrum for each slit through the effective areas of the MUSE channels, apply thermal and instrumental broadening, and place the emission in the detector pixels. We call the resulting spectrogram for this unit of plasma the response function, which is a probability per second, per pixel, and per unit emission measure, that a photon is detected.

For the same unit of plasma in the FOV of another slit, the resulting spectrogram will be similar, but displaced in the spectral direction. If the Doppler velocity of the unit plasma changed, the spectrogram would be shifted by a different amount.  If the temperature of the unit of plasma changed, the overall shape of the spectrogram, and the corresponding detector signal would change. To perform the disambiguation, we compute the response functions
for different combinations of plasma temperature, Doppler velocity and originating slit number. The detector responses are concatenated into a response matrix $\mathcal{R}$. A more detailed description of the response functions is given in Appendix~\ref{sec:sdc_response}.

A spectrogram $\vec{y}$ consisting of contributions from plasma at multiple temperatures, velocities and from different slits is a linear combination of the response functions, i.e.,

\begin{equation}
\vec{y} = \mathcal{R}\vec{x},\label{eqn:system}
\end{equation}

\noindent where components of $\vec{x}$ correspond to the amount of emission measure of plasma for certain combinations of the physical parameters (Doppler velocity, temperature, slit number).  Given an observed spectrogram $\vec{y}$, solving Eq.~(\ref{eqn:system}) for $\vec{x}$ corresponds to solving for a DEM distribution as a function of $v_{\rm Doppler}$, $\log{T}$ and slit number, which we name a VDEMS distribution.

Each of the three MUSE spectral passbands is spanned by $1024$ pixels (e.g. see Figure~\ref{fig:fig3}).  Using the data from all three passbands implies
the input to the SDC is a column vector $\vec{y}$ with $M=3072 = 3 \times 1024$ components.  The number of VDEMS components (i.e. components of $\vec{x}$) $N$ is equal to the product of the bins in $v_{\rm Doppler}$, $\log{T}$ and the number of MUSE slits (37). For applications to MUSE data, $N\sim 10M$, so Equation~(\ref{eqn:system}) is under-determined. We note that the CI 195\AA\ intensities can also be used by the SDC to better constrain the mid-temperature corona, if included, $\vec{y}$ would be larger. When 195\AA\ intensities are included, we assume a response function that is constant in velocity. Including 195 \AA\ intensities improves the constraints on the inversions in the temperature range covered by \ion{Fe}{12}~195\AA.  

To solve Equation~(\ref{eqn:system}), we seek a sparse solution  to minimize the amount of emission measure needed to explain the detector signal $\vec{y}$. This is done by the following minimization problem:

\begin{equation}
\vec{x}^{\#} =  {\rm argmin} \left\{ \frac{1}{2} [\mathcal{R}\vec{x} - \vec{y}]^2 + \alpha|\vec{x}|_{1}\right\},\, x_i  \ge  0 {\rm~for~ i=1,...,N},
\end{equation}

\noindent where $|\vec{x}|_{1}$ is the L1-norm of $\vec{x}$. The regularization parameter $\alpha$ is a hyperparameter that influences the degree of sparsity in the solution. The basis used for VDEMS inversions are normalized top-hat functions that are zero except in specific, individual, bins in ($\log{T}$, {$v_{\rm Doppler}$}, slit number) space (like discrete Dirac delta functions). With the choice of this basis, the posed minimization problem translates into the statement that we prefer a solution that requires the least amount of total emission measure (EM) to explain the spectrogram. Why did we choose this basis? From missions like IRIS, Hi-C, and SDO/AIA (and from 3D MHD models like Bifrost and MURaM), we know that, at spatial scales of 1\arcsec\ or less, emission detected in any spatial pixel can have line-of-sight contributions from distinct loops (with different properties) that happen to cross through said pixel. For the same 3D coronal structure, one perspective may lead to one pixel exhibiting plasma from a $1$ MK loop and from a $7$ MK loop. For a different viewing angle, the pixel may have emission from only the $1$ MK loop or the $7$ MK loop. A similar argument can be made for distributions across velocity space (and certainly in slit number-space).

To solve this system, we use the Lasso Least Angle Regression (implemented as the {\tt LassoLars}) routine in the Python scikit-learn package~\citep{scikit-learn}.  In our validation experiments of the SDC, we tested a range of alpha values (between $10^{-5}$ and $0.5$). Based on these tests, $\alpha=10^{-3}$ provided inversions such that: (a) $[\mathcal{R}\vec{x} - \vec{y}]^2 \sim 1$ (i.e. predicted detector spectrogram is consistent with the synthetic observed spectrogram), and (b) the reconstructed VDEMS is close to the ground truth, such that the dominant line(s) are correctly identified.

\subsection{Using the SDC to calculate main lines \label{sec:sdc_target}}

In this section, we use the SDC code to calculate the main line intensities and velocities.  In Section~\ref{sec:sdc_target_demo}, we provide an example of this method and in Section~\ref{sec:sdc_target_valid} we discuss the many tests that were used to validate this method.  We have validated this code using numerical simulations of waves \citep{Antolin2017}, flux emergence (model B), a quiescent active region (Model A) and a flare (Model C). Here we show results for Models A and C. The conclusions are similar for all models. 

There are two shortcomings to the  quiescent active region model A.  The first is that the maximum temperature in the active region is $\sim$ 3.5 MK.  This implies that there will be no measurable \ion{Fe}{19} 108\AA\ emission in the MUSE spectra of this simulation.  Because of this, we also include an example of the  \ion{Fe}{19} 108\AA\ emission from the flare simulation (model C).

Secondly, the transition region in model  A is artificially broadened to reduce the resolution requirements in the transition region \citep{lionello09}.  Though this has been shown to not impact the coronal solution \citep{mikic13}, this assumption increases the transition region emission measure to substantially more than is observed.  This implies we can only use emission measure of the simulation from temperatures above 0.5~MK in this analysis.  We have extended the emission measure solution to lower temperatures in two ways.  First, we simply use the standard CHIANTI active region DEM and extrapolate from the lowest temperature bin at each spatial location.  Second, because we want to investigate the potential impact of the \ion{He}{2}~304\AA\ line in the 284\AA\ channel, we use the \ion{He}{2} 304\AA\ channel intensities from Active Region 7986 observed by the Extreme-ultraviolet Imaging Telescope (EIT, \citealt{eit_instr}) and the EIT calibration \citep{dere2000} to estimate the emission measure at Log T = 5.0.  This estimate implies that the \ion{He}{2} intensities we will forward calculate in the MUSE data are consistent with the observed \ion{He}{2} intensities from the original active region observation.

\subsubsection{An Example of using the SDC}

\label{sec:sdc_target_demo}
Each simulation provides a cube of emission measure as a function of temperature and velocity at all spatial locations in the active region or flare; this is the true VDEM cube. We then take those emission measure cubes and simulate the MUSE spectra and context images, see Figure~\ref{fig:fig3} for an example of synthetic MUSE data.  
The 37 MUSE slits will only sample part of the emission measure cube at a single pointing, so MUSE will raster over the 4.45\arcsec\ distance between slits to build up the spatial information for the entire active region. That is, we simulate a series of detector images, each for a different pointing.  When we invert them, we can build up an inverted 3D VDEMS cube for all spatial locations (along the slit, and for each raster step position).  

When we simulate the MUSE data, we use response functions calculated with a specific set of CHIANTI input parameters, for instance, a pressure of 
$3\times 10^{15}$ cm$^{-3}$ K and coronal abundances  (for more information on the response functions, see Appendix~\ref{sec:sdc_response}).  We 
use an exposure time of 1\,s for the AR simulation and 1.5\,s for the flare simulation to add photon noise using the IDL Poisson random number generator for the detector images.

From the true and inverted VDEM cubes, we calculate spectrally pure intensities (again assuming the same pressure and abundances) and velocities for the three main lines, namely \ion{Fe}{9} 171\AA, \ion{Fe}{15} 284\AA, and \ion{Fe}{19} 108\AA.  The intensities and velocities from the true VDEM cubes become the ``ground-truth'' to which we compare the inverted data.  

\begin{figure*}
    \centering
    \includegraphics[width=0.7\textwidth]{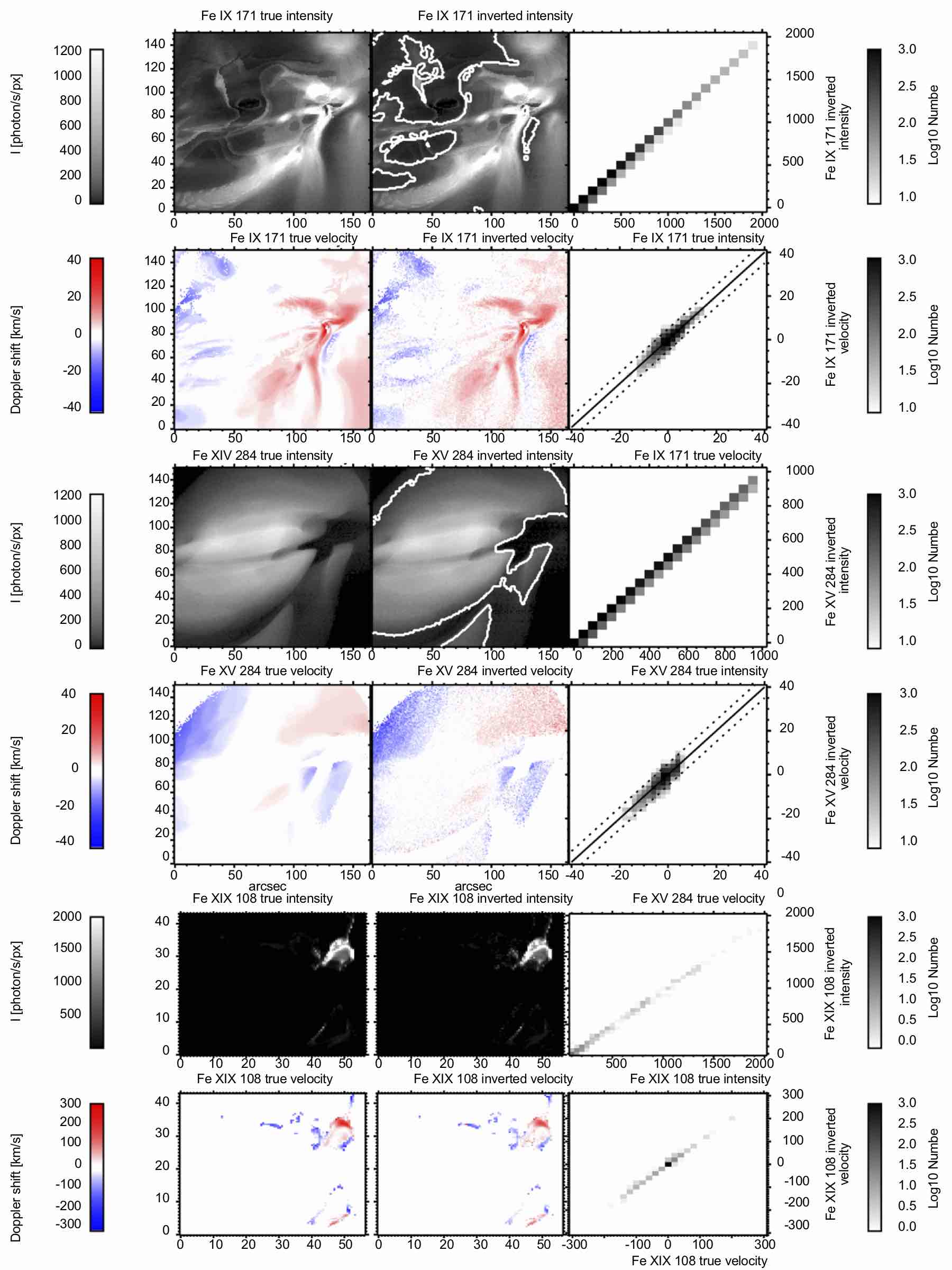}
    \caption{Illustration of our novel Spectral Disambiguation Code (SDC) that successfully resolves multi-slit confusion in the MUSE spectra. Comparison of ground truth (left column) \ion{Fe}{9}~171\AA,  \ion{Fe}{15}~284\AA, \ion{Fe}{19}~108\AA\ intensities (even rows) and velocities (odd arrows) from numerical simulations of an active region (mdel A, top 4 rows) and a flare (model C, bottom two rows) with intensities and velocities from the SDC inversion (middle column) show very good agreement. Joint probability density plots (right column) compare the true and inverted values; for velocities the plot only shows locations with $S/N > 10$, illustrated with the contoured intensities. The true signal does not include photon noise, while the SDC inversion is based on inverting MUSE data that includes photon (Poisson) noise for a 1~s and 1.5~s exposure time for the quiescent AR and the flare simulations, respectively. Dotted lines show $\pm 5$ 
    km~s$^{-1}$ error bars for 171 and 284\AA.}
    \label{fig:fig12}
\end{figure*}

Figure~\ref{fig:fig12} shows an example of this comparison.  
The left column shows the ground truth for the total intensity and first moment for the active region simulation in the \ion{Fe}{9} 171\AA\ line and \ion{Fe}{15} 284\AA\ line and the flare simulation in \ion{Fe}{19} 108\AA\ line.  The second column shows the same moments calculated from the VDEM cube inverted from the simulated, noisy MUSE data.  In the inversion, we used response functions assuming the same pressure and abundances that generated the original data.  
 The contours are the locations where the intensity is larger than 100 photons s$^{-1}$ exposure$^{-1}$ (S/N=10). A comparison of the intensities and first moments is shown in the final column (for pixels with S/N$>$ 10).  These results demonstrate that the characteristics of the primary spectral lines can be well determined using the SDC. 

\subsubsection{Validation tests}
\label{sec:sdc_target_valid}

We then ran a series of tests to determine how well the SDC returned the main line intensities and velocities with different inversion parameters.  We summarize the inversion parameters considered in Table~\ref{tab:invparam}.    We discuss the impact of each of these inversion parameters briefly below.  In all cases, we relate the ability to invert the data to the example given in Section~\ref{sec:sdc_target_demo}.   For this case, the abundance and pressure used to calculate the inversion response functions matched the true abundance and pressure, namely a pressure of 
3$\times$ 10$^{15}$ cm$^{-3}$ K and coronal abundances.  The temperature and velocity ranges encompassed the entire temperature and velocity ranges in the simulations.  The results given in Section~\ref{sec:sdc_target_demo}, then, represent the best case scenario for the inversion.  Poisson noise dominates the uncertainty in the intensities and velocities.  

\begin{deluxetable}{c|c|c}
\tablecaption{Summary of Inversion Parameters \label{tab:invparam}}
\tablehead{
\colhead{Parameter} & \colhead{Active Region} & \colhead{Flare} \\
& \colhead{Values Considered} & \colhead{Values Considered}
}
\startdata
Abundances &\multicolumn{2}{c}{Coronal or Photospheric} \\
Pressure & \multicolumn{2}{c}{3$\times$ 10$^{14}$,  3$\times$ 10$^{15}$, or 3$\times$ 10$^{16}$ cm$^{-3}$ K }\\ 
Spectral Windows& \multicolumn{2}{c}{All combinations and 195 Included or Not Included}\\
Temperature Range & Log T = 4.6 - 6.6 or 5.6 - 6.6 &   Log T = 4.7 - 7.5\\
Temperature Resolution & $\Delta$ Log T = 0.1 or 0.2 &  $\Delta$ Log T = 0.2\\
Velocity Range & $\pm 20$ km s$^{-1}$ or $\pm 50$ km s$^{-1}$  & $\pm 400$ km s$^{-1}$ \\
Velocity Resolution & 5 km s$^{-1}$ & 20 km s$^{-1}$\\
\enddata
\end{deluxetable}

\begin{figure*}
    \centering
    \includegraphics[width=0.7\textwidth]{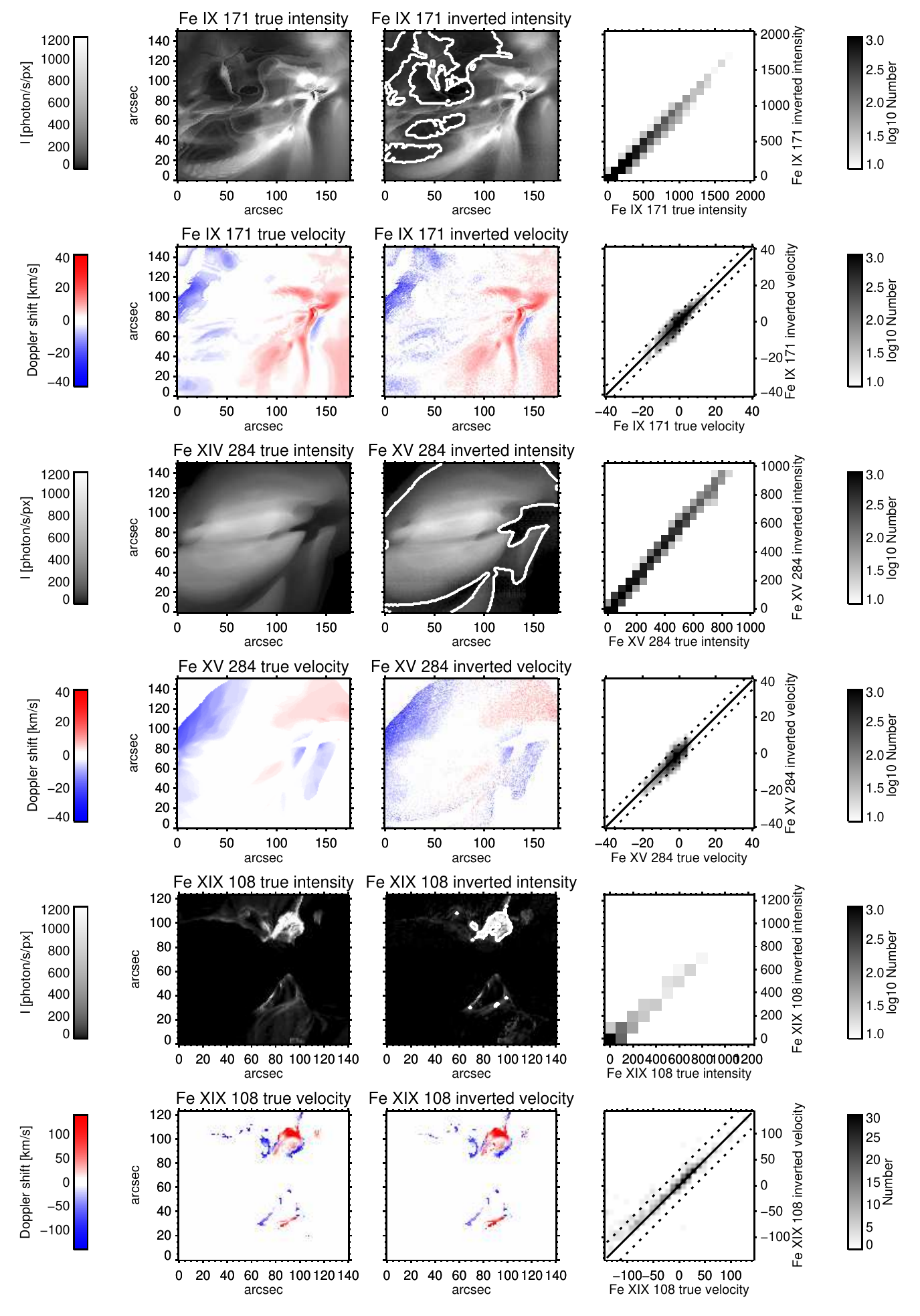}
    \caption{Same as Figure~\ref{fig:fig12} but the response functions used in the inversion have different abundances than the ones used to calculate the MUSE intensities.  The only difference between this result and the one shown in Figure~\ref{fig:fig12} is the somewhat broader distribution of intensities in the main lines.}
    \label{fig:moments_photo}
\end{figure*}

First, we inverted the data using response functions calculated with abundances different than the abundances used to calculate the detector data, meaning, for instance, the detector data was calculated with coronal response functions, then inverted with photospheric response functions.  There was no impact on the ability to reproduce the velocities or widths when using these different response functions; the ability to determine these parameters was identical to the example given in Section~\ref{sec:sdc_target_demo}. The only effect was a somewhat broader distribution in the absolute intensity (Figure~\ref{fig:moments_photo}).
Though the primary lines in each channel are from different ionization stages of Fe, the contaminant lines are not all Fe.  The inversion attempts to find an emission measure that can recreate both main and contaminant spectral lines, but because the response function is generated with different abundances, it finds an emission measure that minimizes the difference.  This causes a small additional uncertainty in the intensity of the main lines.  Compared to all other inversion parameters, discussed below, using different abundances in the inversion had the largest impact. We note that this problem does not affect the main lines in our baseline approach, in which the SDC is used only to estimate the contaminant lines (\S~\ref{sec:sdc_baseline}). In addition, a future expansion of the SDC code can easily incorporate abundance variations as an additional parameter to invert. 

Next, we determine the impact of using, for the inversion, a pressure that is different from the pressure used to calculate the (forward) synthetic data.  We considered  pressures of 
3$\times$ 10$^{14}$ cm$^{-3}$ K and 3$\times$ 10$^{16}$ cm$^{-3}$ K.  We found no difference in the ability to determine the intensities, velocities and widths with the incorrect pressure.  This is due to the fact that the main lines are only weakly dependent on electron pressure. We conclude that inverting the data without knowledge of the true pressure of the plasma will have no real impact on the results.

We also performed the inversion with and without the 195\AA\ context imager data.  When we included the 195\AA\ intensities, the inverted emission measure cube better predicted the 195\AA\ intensity than when we did not. 

Finally, we ran a series of tests to determine the impact of the temperature range and resolution and velocity range on the inversion.  For the active region simulation, the temperature ranges considered included $\log (T[K]) = 4.5 - 6.5$ or $\log (T[K]) = 5.5 - 6.5$.  The first temperature range includes the potential impact of the \ion{He}{2} in the response, the second does not.  We find both temperature ranges can adequately predict the intensities and velocities of the \ion{Fe}{9} and \ion{Fe}{15} lines, meaning even though the ground truth data was calculated with a realistic \ion{He}{2} contribution, ignoring it did not change the solution.   Likewise the choice of velocity range did not impact the results.

\subsection{Using the SDC to subtract contaminants} \label{sec:sdc_baseline}

\begin{figure*}
    \centering
    \includegraphics[width=0.5\textwidth]{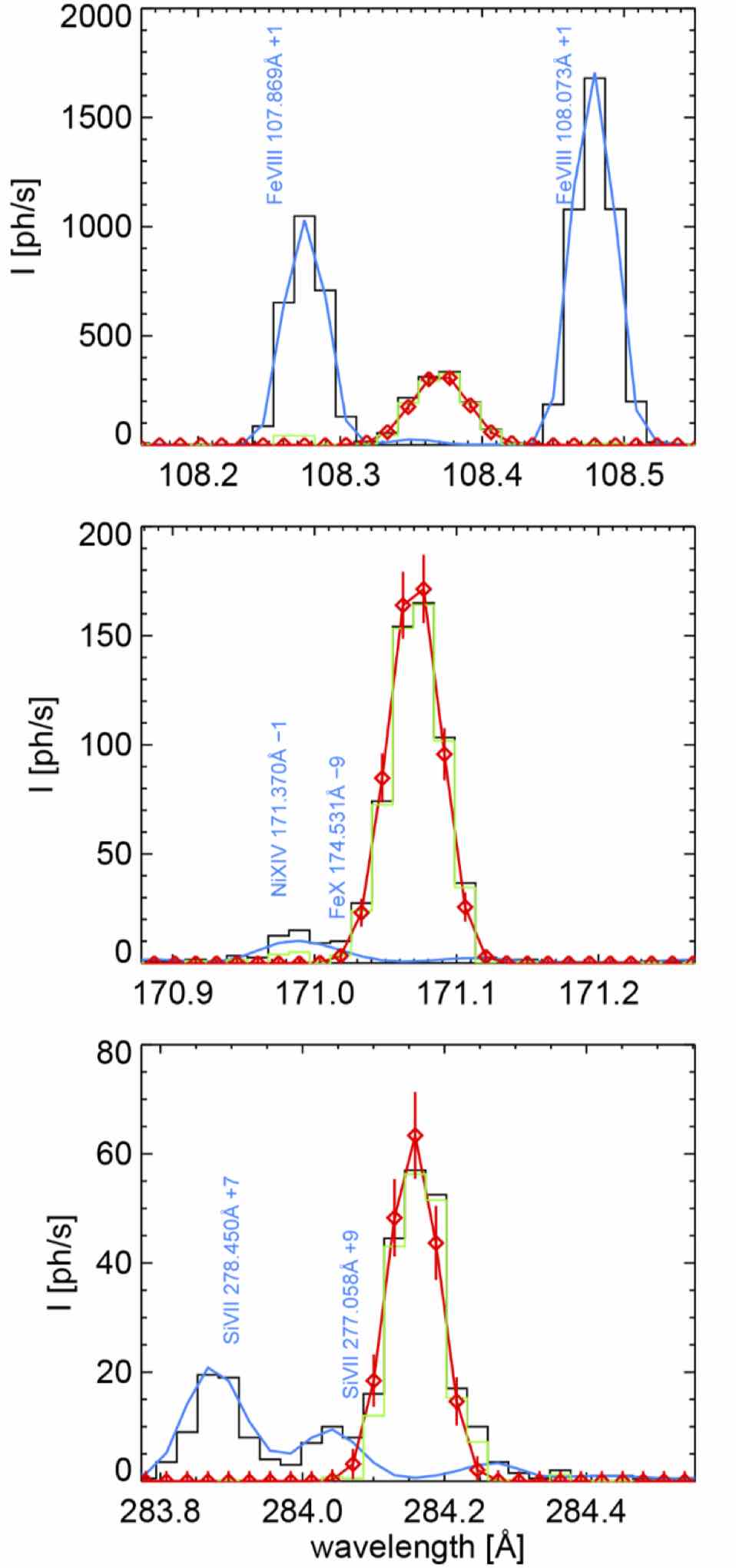}
    \caption{{Examples of using the SDC to subtract contaminant lines. In locations where secondary lines from neighboring slits impact the main line (ground truth with Poisson noise in red, from model C (top row) and model B (middle and bottom rows)), SDC inversion results can be used to remove the effects of these secondary lines (blue) from the MUSE signal (black) and isolate the estimated main line (green). The main differences between the ground truth (red) and derived main line are due to photon noise.}}
    \label{fig:decontam}
\end{figure*}

In the previous sections we described our efforts to quantify the accuracy of the SDC in reproducing the properties of the main lines. Here we discuss an alternative use of the SDC analysis, to subtract the multi-slit contamination from the MUSE spectra to allow an analysis of the ``de-contaminated" spectra. This is the approach we have baselined for MUSE.
The VDEMS obtained from the SDC analysis can be used to synthesize the MUSE spectra of the contaminant only (i.e., using response functions including all contributions except the main lines) in the three spectral windows. We note that the line profiles of the contaminants are not necessarily Gaussian for two reasons: the shape of the response functions is not Gaussian (Appendix~\ref{sec:sdc_response}) and although thermal broadening is Gaussian, the velocity distribution in VDEM space is typically not.

\begin{figure*}
    \centering
    \includegraphics[width=0.95\textwidth]{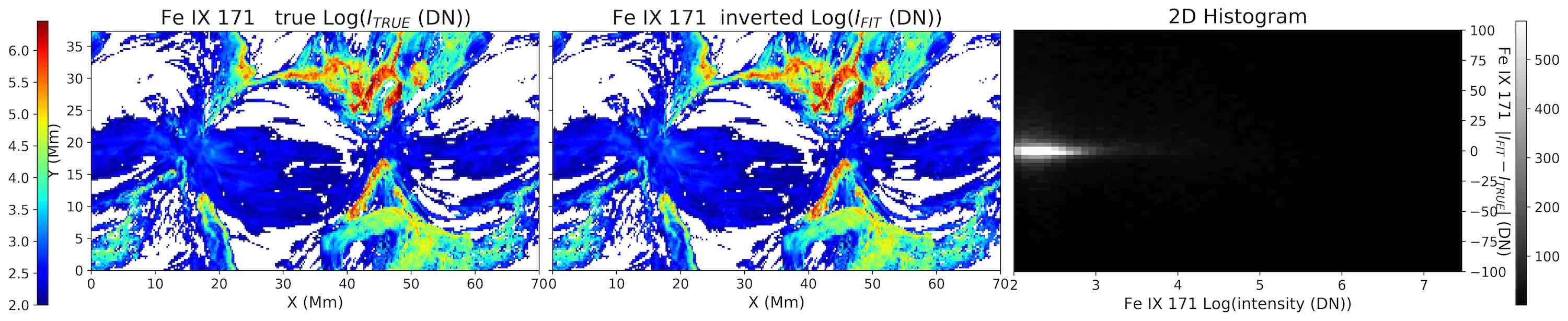}
    \includegraphics[width=0.95\textwidth]{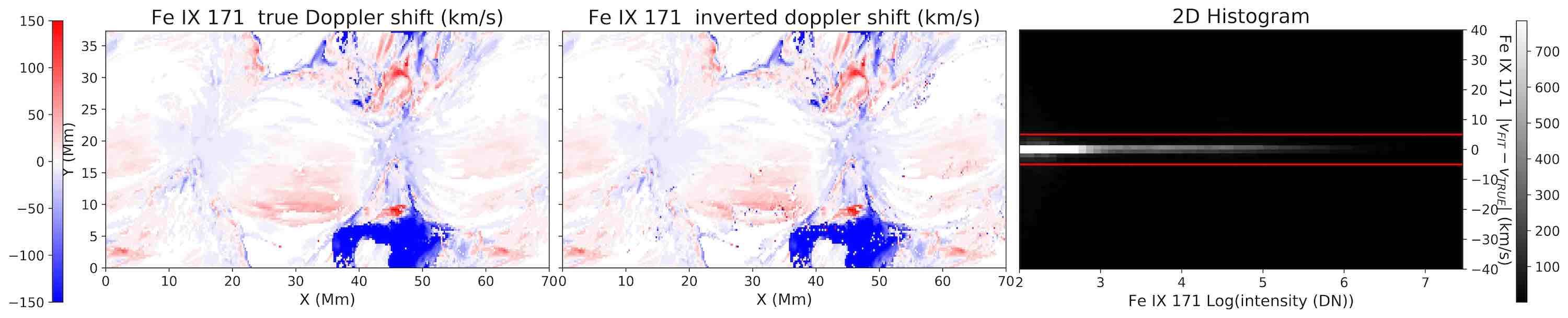}
    \includegraphics[width=0.95\textwidth]{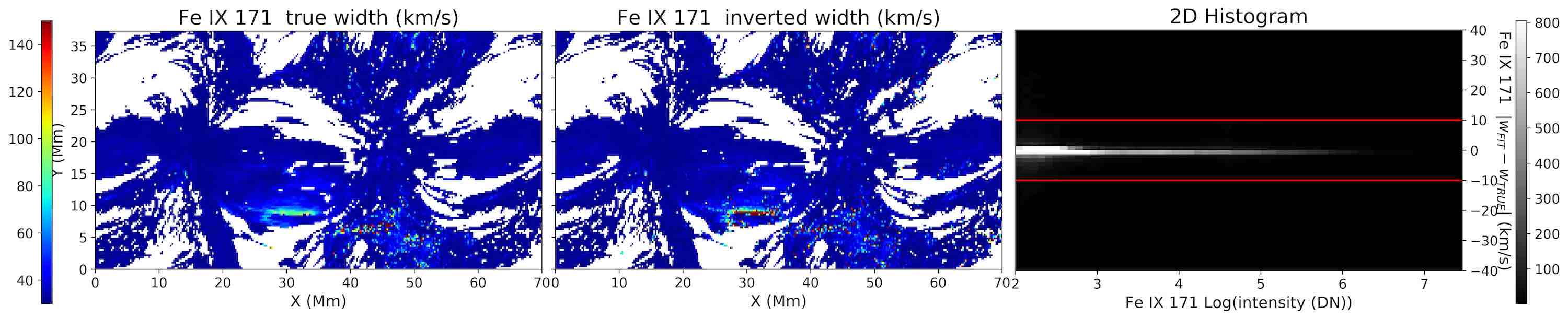}    
    \caption{The accuracy in determining the main line parameters when using the SDC to subtract contaminants. Left column: Ground truth total intensity (top), Doppler shift (middle), line width (bottom) of the main spectral line in the 171\AA\ passband as computed from the true VDEMS (i.e., the VDEMS derived from the physical parameters -- density, temperature, velocity, slit position -- in the simulation).
    Middle column: the corresponding parameters derived from single Gaussian fits to the profiles in which the contaminants are determined using the SDC and subtracted from the MUSE signal. Right column: JPDF of the relative difference between the ground truth line properties and those computed from the simulated MUSE spectra after subtraction of contaminants, as a function of intensity. 
    To guide the eye, the red lines in the middle-right panel show $\pm 5$~km~s$^{-1}$ uncertainties and $\pm 10$~km~s$^{-1}$ uncertainties for the bottom-right panel . 
    }
    \label{fig:sub_gausfit_171}
   \end{figure*}
  
Figure~\ref{fig:decontam} shows an example of this approach, applied to spectra calculated from the flare simulation of model C (for the \ion{Fe}{19} 108\AA\ line; top), and the flux emergence simulation of model B (for the \ion{Fe}{9} 171\AA, and the \ion{Fe}{15} 284\AA\ lines; middle and bottom). We have intentionally chosen locations where the contamination from other slits is significant. This is often when the main line is not as bright. Black lines show the full simulated MUSE spectrum, while blue lines arise from the SDC inversion for contaminant lines. The blue lines are used to subtract the contaminants from the full MUSE spectrum. The de-contaminated spectra (green) are very close to the ground truth (red): the differences are within the photon noise. 

 \begin{figure*}
   \centering
    \includegraphics[width=0.95\textwidth]{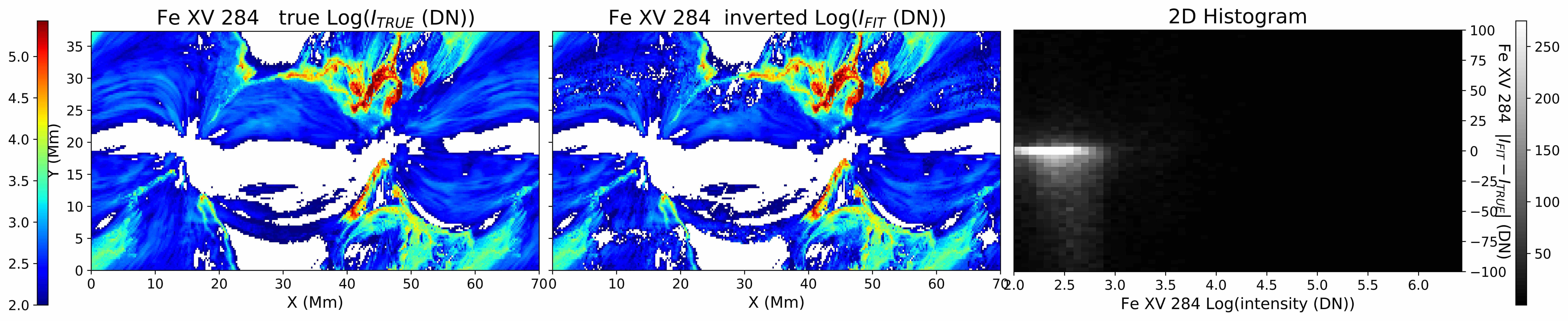}
    \includegraphics[width=0.95\textwidth]{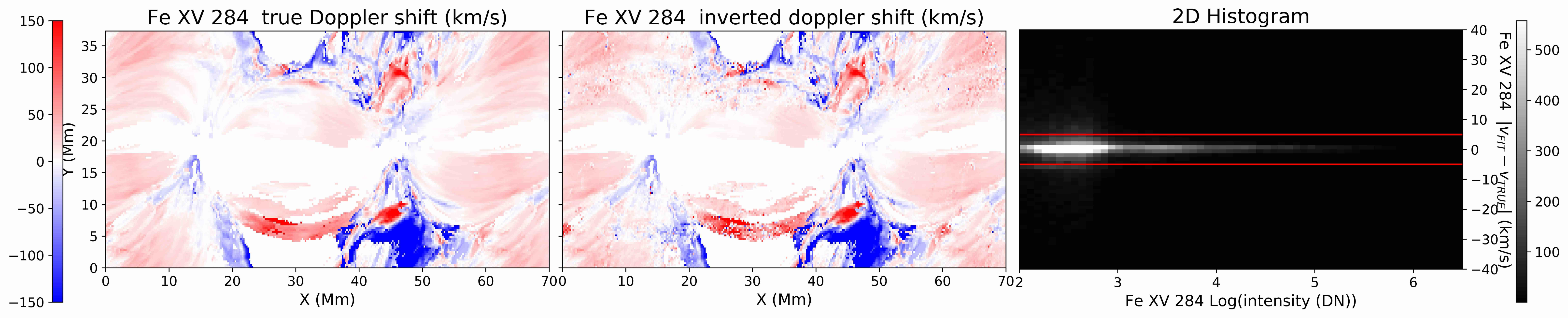}
    \includegraphics[width=0.95\textwidth]{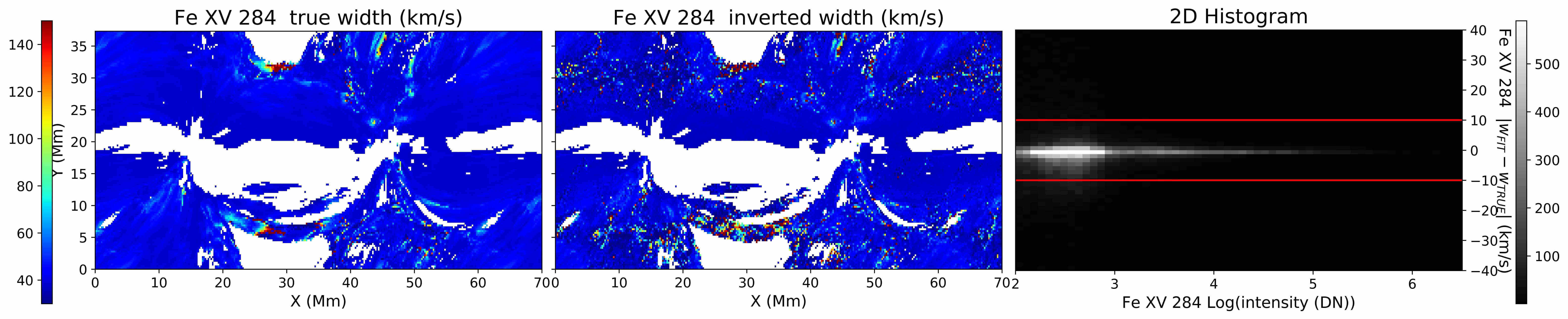}
    \caption{As in figure~\ref{fig:sub_gausfit_171} but for the 284\AA\ passband.}
    \label{fig:sub_gausfit_284}
    \end{figure*}

\begin{figure*}
    \centering
    \includegraphics[width=0.95\textwidth]{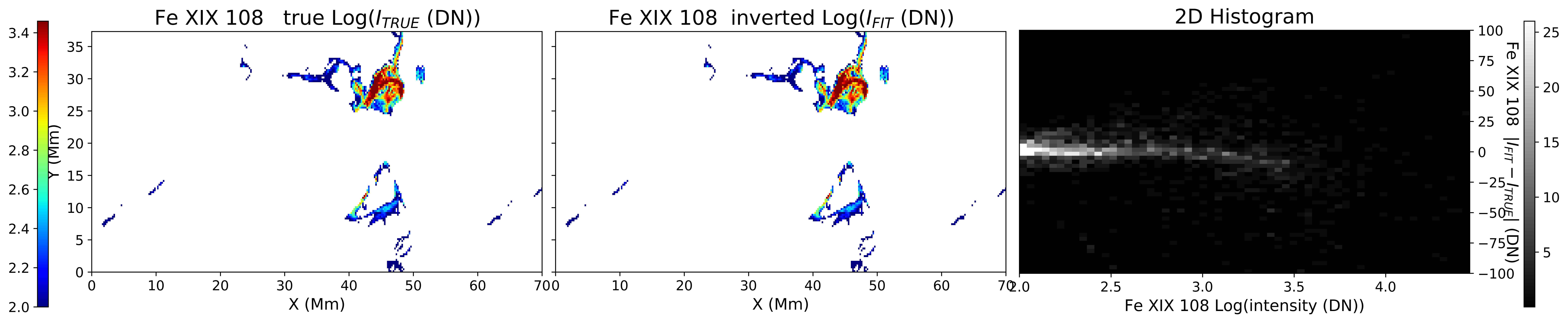}
    \includegraphics[width=0.95\textwidth]{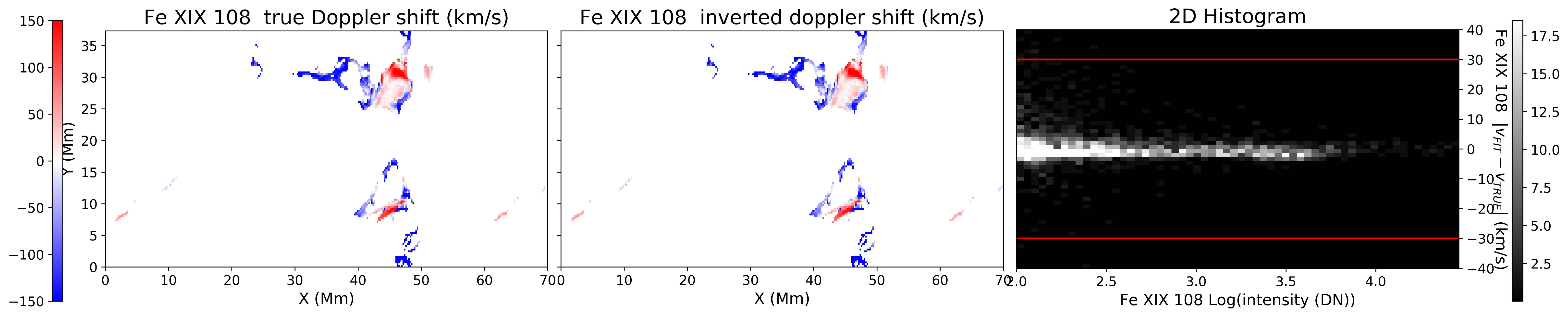}
    \includegraphics[width=0.95\textwidth]{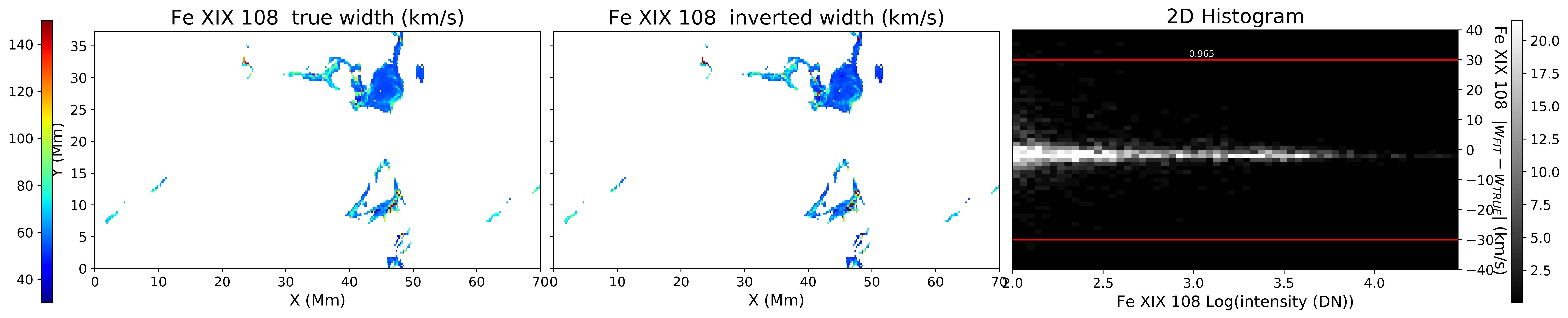} 
    \caption{As in figure~\ref{fig:sub_gausfit_171} but for the 108\AA\ passband.
    The red lines show $\pm 30$~km~s$^{-1}$ uncertainties.
    }
    \label{fig:sub_gausfit_108}
\end{figure*}

In Figures~\ref{fig:sub_gausfit_171}-\ref{fig:sub_gausfit_108} we show the application of this approach to the flare simulation of model C for all three spectral bands. We calculate simulated spectra, include Poisson noise, and then apply the SDC. We use the VDEMS resulting from the SDC analysis to subtract the contaminants from the simulated spectra, and fit the de-contaminated spectra of the main lines with single Gaussians to derive the intensity, Doppler velocity and line width. Figures~\ref{fig:sub_gausfit_171}-\ref{fig:sub_gausfit_108} compare total intensity (top row), Doppler shift (middle row) and line width (bottom row) of the ground truth of the main line (left column) with the corresponding parameters determined from the decontaminated MUSE data  (middle column) and Joint Probability Distribution functions (JPDF) of the relative difference between the ground truth parameters and the parameters determined after subtracting the contaminants
(as a function of total intensity). As mentioned before, this case represents a worst case scenario in terms of potential multi-slit ambiguities: 1) it shows a flare, which typically shows much larger velocity gradients and cool contaminantes than typical ARs; 2) because we tile the FOV of the flare simulation twice in order to fill the MUSE FOV, this implies that MUSE would be observing two flares occurring at the same time within its FOV. 
Even in this absolute worst case, Figures~\ref{fig:sub_gausfit_171}-\ref{fig:sub_gausfit_108} illustrate that the line properties can be inferred accurately within the desired uncertainties (red solid lines in the right column).

\subsubsection{Robustness of SDC contaminant subtraction approach to inversion parameters}

\label{sec:sdc_robustness}
In the previous section we showed that using the SDC to subtract contaminants allows the determination of line parameters to within the desired uncertainty.  Here we perform a similar series of inversions, but by using different pressures, abundances, and spectral windows, calculated with and without noise, we investigate the robustness of our method to the varying assumptions.

\begin{figure*}
    \centering
    \includegraphics[width=0.65\textwidth]{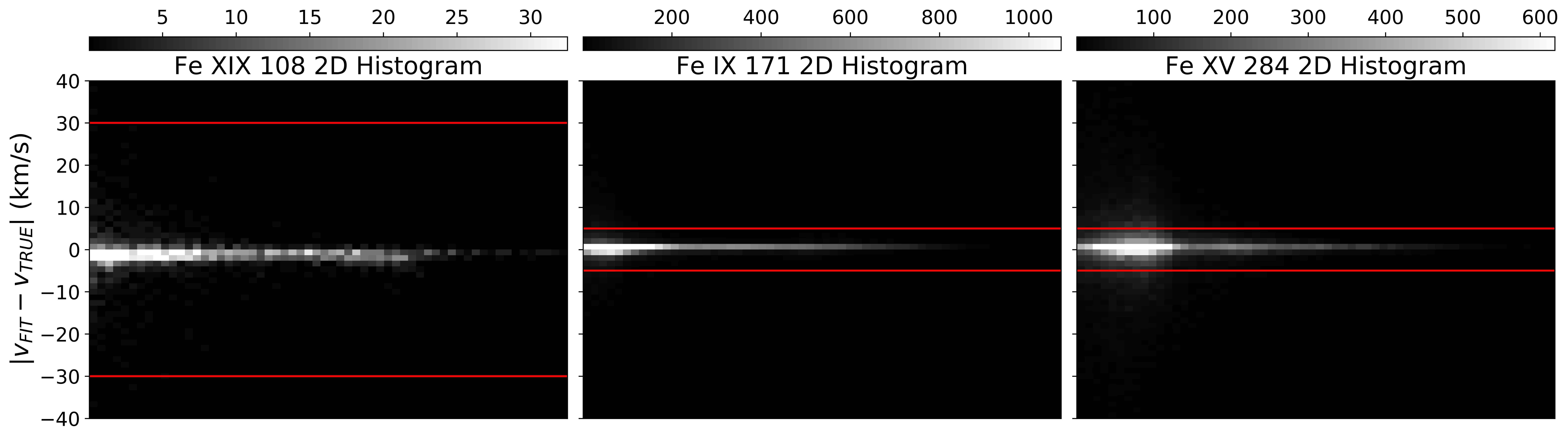}
    \includegraphics[width=0.65\textwidth]{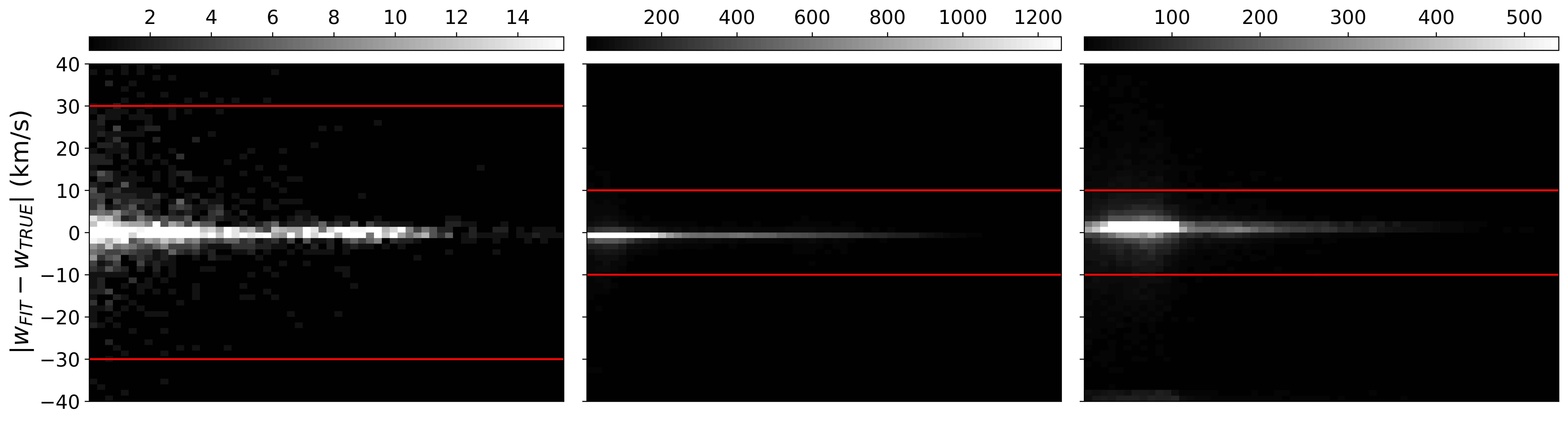}
    \includegraphics[width=0.65\textwidth]{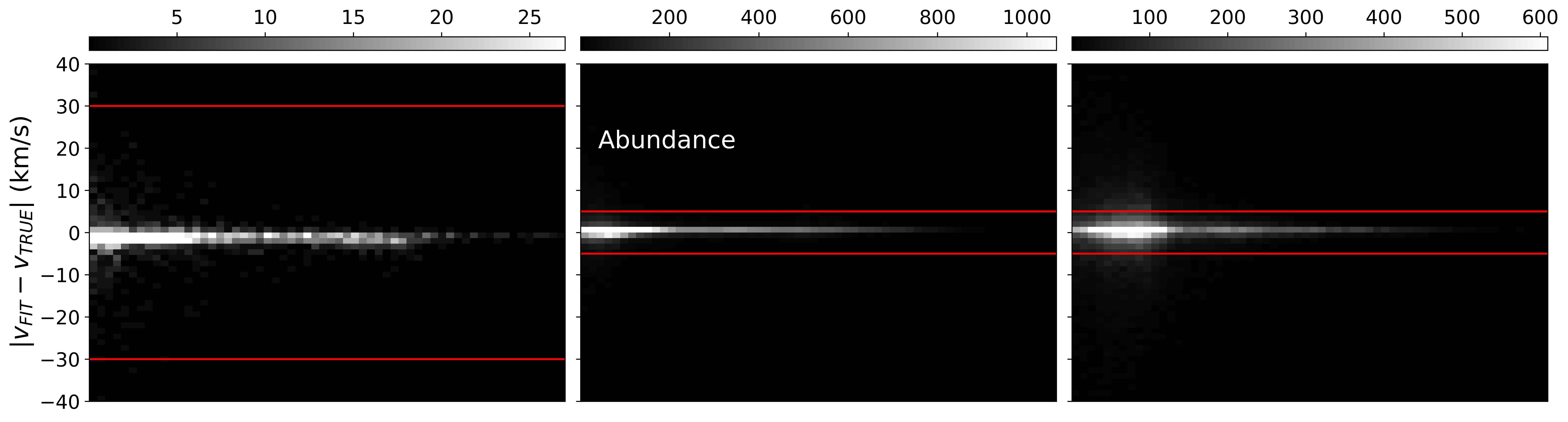}
    \includegraphics[width=0.65\textwidth]{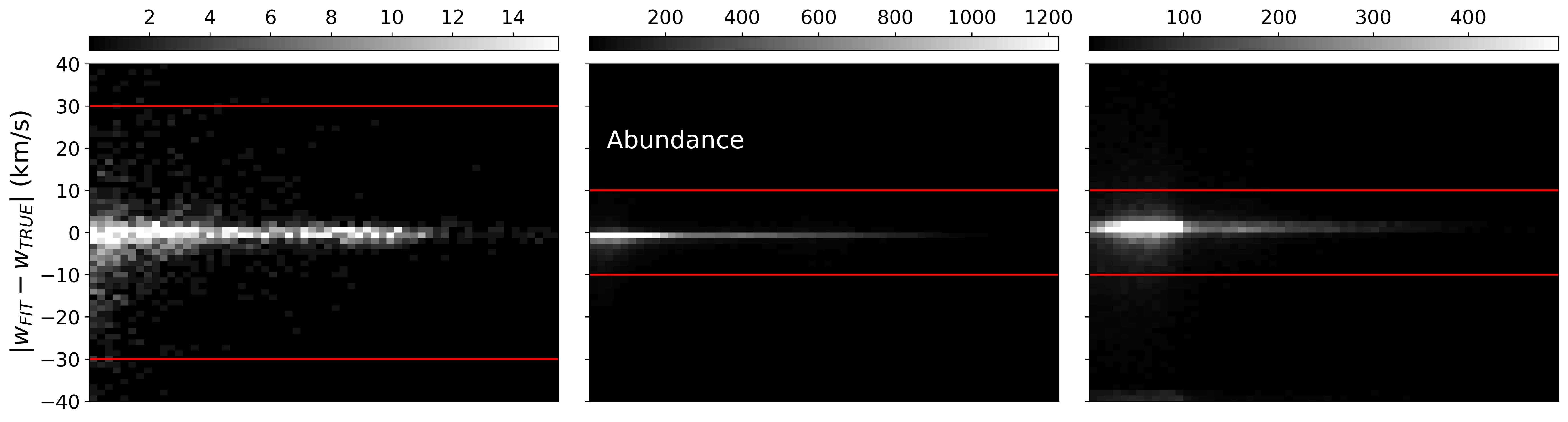}
    \includegraphics[width=0.65\textwidth]{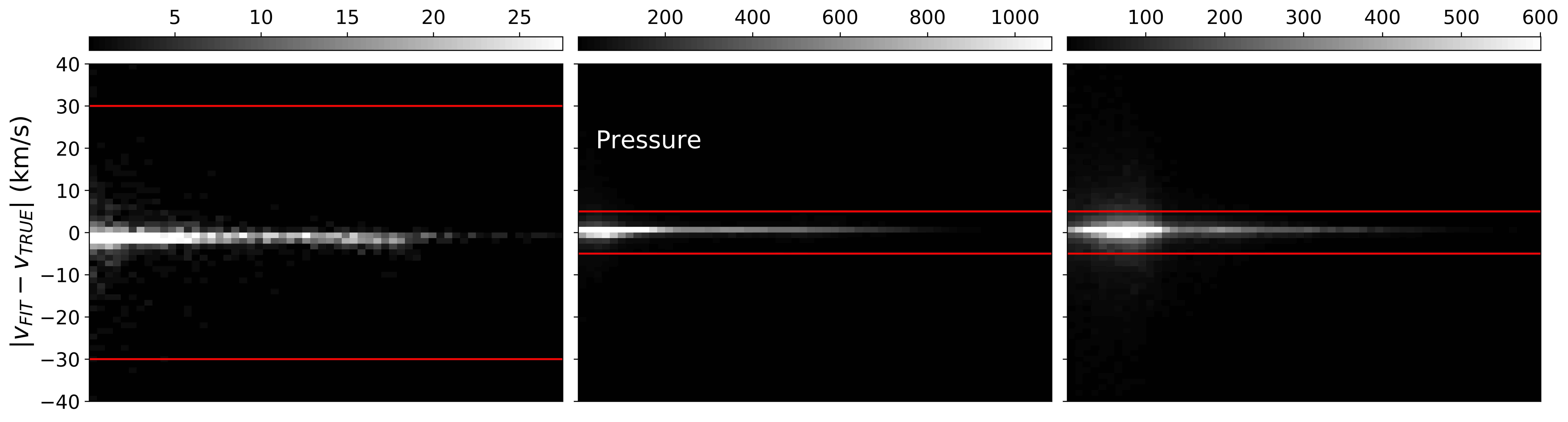}
    \includegraphics[width=0.65\textwidth]{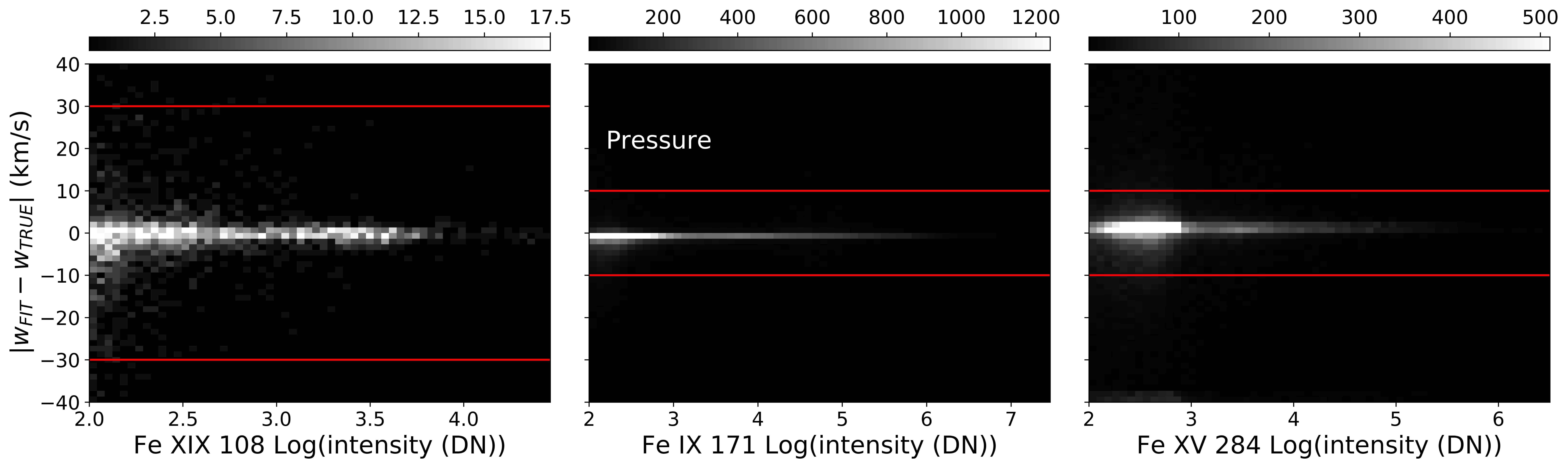}
    \caption{Results from subtracting contaminants that are determined from SDC inversions. Top two rows use the same response function for synthesizing and inverting MUSE data. Middle two rows assume photospheric abundances for synthesis of MUSE data, and coronal abundance for the inversion. Bottom two rows assume electron pressure of $3\times10^{15}$~cm$^{-3}$~K to synthesize MUSE data and electron pressure of $3\times10^{16}$~cm$^{-3}$~K to invert. Each set of two rows shows Doppler shift (top) and line width (bottom). These JPDFs are similar to the right column of Figure~\ref{fig:sub_gausfit_171}. The left, middle, and right column are for \ion{Fe}{19}, \ion{Fe}{15}, and \ion{Fe}{9}, respectively.
    \label{fig:sub_mini}}
\end{figure*}

Figure~\ref{fig:sub_mini} shows JPDFs of the relative errors of Doppler shift and line width for  \ion{Fe}{19}~108\AA, \ion{Fe}{9}~171\AA, and \ion{Fe}{15}~284\AA\ for three different experiments in which the contaminants have been subtracted, as detailed in the previous section. 
\begin{itemize}
\item Experiment 1: The MUSE spectrum is synthesized using the same response function that is used to invert the data (i.e., like in the previous section).
\item Experiment 2: The MUSE spectrum is synthesized assuming photospheric abundances (instead of coronal abundances), while the inversion assumes coronal abundances. 
\item Experiment 3: The MUSE spectrum is synthesized using a response function that assumes a pressure of 
in $3\times 10^{15}$~cm$^{-3}$~K, while the inversion assumes a much larger pressure ($3\times 10^{16}$~cm$^{-3}$~K).
\end{itemize}

All three experiments show that the derived parameters can be determined within the desired uncertainty, and that the differences between the various experiments are negligible. Note that these experiments have been done for a possible worst case scenario, i.e.,  the flare simulation of model C that has been tiled twice to fit within the MUSE FOV, simulating the effects of two simultaneous flares occurring next to each other.

The very good agreement between the ground truth and inverted parameters despite the varying assumptions is caused by several factors:

\begin{itemize}
\item The contamination arising from the multi-slit ambiguity is minimized by the choice of the interslit spacing and the relatively isolated bright lines.

\item The SDC method incorporates a wealth of information from all three spectral windows and the context imager. This information constrains the contaminants well.

\item The spectral lines in our passbands are relatively insensitive to variations in coronal pressure and, to a lesser extent, abundance.

\item {By removing the contaminants instead of synthesizing the main lines, we further minimize the uncertainties in our assumed abundances and pressures. The contaminant lines are usually weak, and subtracting a weak signal that contains a small error, due to either an unknown abundance, pressure, or other artifact, represents a negligible change in our interpretation of the main line.} 
\end{itemize}

In summary, the SDC method requires that assumptions be made on the pressure and abundance of the emitting plasma. The MUSE baseline has been carefully selected to drastically minimize any of these uncertainties. Our results show that such uncertainties are not significant.

\section{Deep Learning Approach}
\label{sec:sdc_dnn}
In addition to the Gaussian centroiding and SDC approaches, the MUSE team are developing a line-fitting approach based on deep neural networks (DNNs). The aim is to extract the zeroth, first, and second moments of individual lines as they appear in individual slits using the MUSE spectrogram (with all three spectral bands) as input to the DNN. 

\begin{figure*}
\centering
\includegraphics[width=\textwidth]{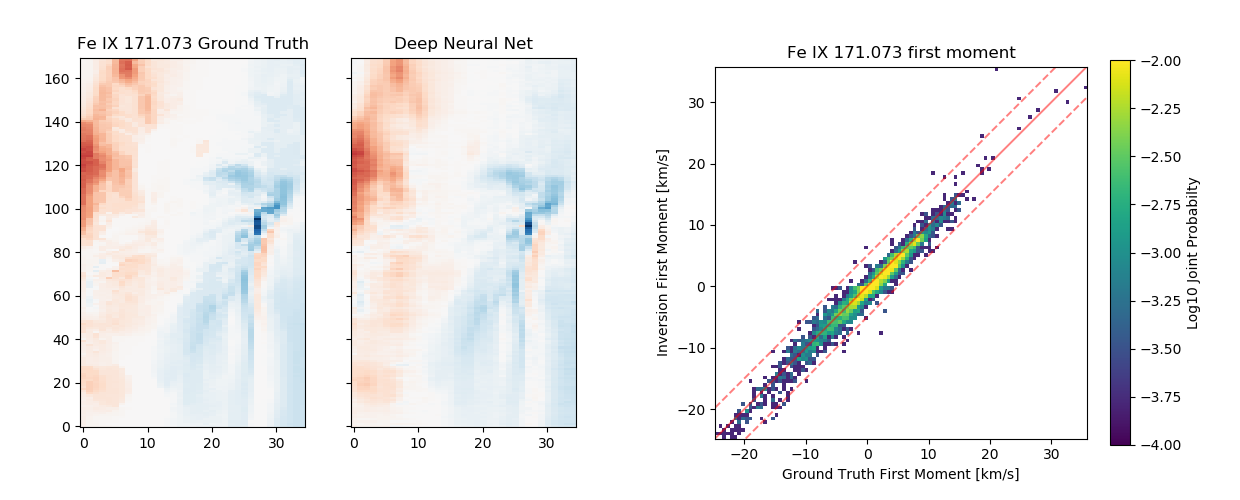}
\caption{Comparison between ground truth (left panel) 1st moment of the~\ion{Fe}{9}~171\AA~line, and the quantity as extracted by a trained deep neural net (DNN) from synthetic MUSE spectrograms. The right panel shows a joint probability density function of the two quantities, and shows how the DNN-extract 1st moment 
satisfies the $5$ km s$^{-1}$ desired uncertainty.}\label{fig:dnn_fe_9_first}
\end{figure*}

Figure~\ref{fig:dnn_fe_9_first} shows an example of the first moment of the \ion{Fe}{9}~171\AA~line from model A of the quiescent AR. The left panel shows the ground truth as computed from the simulation, and the middle panel shows the quantity as extracted by a trained DNN. The DNN was trained using data from the same simulation, but for other raster positions. The right panel shows the JPDF between the ground truth and extracted 1st moments. If the inversion were perfect, the JPDF would be the solid red line along the diagonal. There are some deviations, but, as indicated by the red dashed lines, more than $99\%$ of the inverted values lie within $\pm 5$ km s$^{-1}$ of the ground truth. Based on preliminary tests with this MHD model, the DNN approach works similarly well for the zeroth and second moments for this line, and moments of the other dominant lines in the MUSE passband. The DNN approach is not required to satisfy the desired maximum MUSE measurement uncertainty. Nevertheless, in the future we will study the DNN approach in depth to establish whether this machine learning approach can be used for MUSE data pipeline processing.

\section{Discussion} \label{sec:discussion}

The MUSE multi-slit spectrograph is a radically innovative instrument that will for the first time ``freeze" solar evolution and reveal previously invisible processes. It will revolutionize our view of the physical processes that drive coronal heating and the flares and eruptions that lead to space weather.  We have described the multi-slit approach of the MUSE multi-slit spectrograph. We have shown that through a careful choice of main passbands that contain bright and relatively isolated lines, and interslit-spacing, the overlap between signals from neighboring slits that is inherent to a multi-slit approach, is minimized. It is reduced sufficiently to allow determination, in regions where the lines are bright, of the Doppler shift and line width of the main lines (\ion{Fe}{19} 108\AA, \ion{Fe}{9} 171\AA, and \ion{Fe}{15} 284\AA) to within, respectively 30, 5, and 5 km~s$^{-1}$ (Doppler shift), and 30, 10, and 10 km~s$^{-1}$ (line width), as desired.

Through detailed studies of synthetic spectra from advanced numerical simulations, we show that in many locations on the Sun this can be achieved through simple Gaussian fitting without further consideration of multi-slit ambiguity. In some locations (e.g., strong TR emission from cool regions) and under some conditions (e.g., eruptions), multi-slit ambiguity may be present and require a different approach. The baseline approach of the MUSE project will be to automatically flag those locations in the data pipeline (as a level 2.5 data product), and provide the community with the tools to successfully disambiguate the data. This approach is based on a Spectral Disambiguation Code (SDC), first developed by \citet{Cheung:Multicomponent} for MUSE. The automated level 2.5 data product that flags locations that may be subject to multi-slit ambiguity is based on running the SDC inversion at lower resolution. We described the extensive testing we have performed of the SDC and how it can be used to determine the main line parameters to within the desired uncertainty.  The various tests included:
\begin{itemize}
    \item  different Log T and velocity resolution
	\item  different ranges of Log T and velocity
	\item with/without MUSE 195 images included and/or all kind of possible combinations of the various spectral windows
	\item  different pressures for synthesis than for inversion
	\item  different abundances for synthesis than for inversion
	\item  different inter-slit spacing (as part of our optimization study of the inter-slit spacing)
\end{itemize}

We have found that the SDC code is very robust against uncertainties in the
parameters assumed for the inversion
(abundance variations, pressure differences, poorly known lines, e.g., \ion{He}{2} 304\AA). These uncertainties are below those that are caused by photon noise. This is fundamentally because the contaminant lines are much fainter than the main lines.  

This helps both of the approaches we have highlighted in this paper. In the first approach, one could use the SDC to directly determine the main line parameters. The second approach is based on using the SDC to determine the contaminants, subtract those from the MUSE signals, and then proceed to analyze the MUSE signal to determine the main line parameters. We have shown that both approaches work well and can be used to accurately determine physical parameters in the solar atmosphere. Preliminary analysis suggests that subtracting the contaminants might be less sensitive to 
inversion parameter assumptions, as it leaves the main MUSE signal intact, and any uncertainties only affect the faint contaminants. One can imagine also using the SDC to simply identify contaminants and use those identifications as initial parameters to guide a multiple Gaussian fit to the MUSE spectra. This would ultimately be up to the end user. With the advent of machine learning and artificial intelligence techniques, it is highly likely that the SDC itself might be superseded by deep neural network approaches that build on the preliminary approach highlighted in \S~\ref{sec:sdc_dnn}. 

\acknowledgments
This paper is dedicated to the memory of Ted Tarbell, who led the MUSE team from its inception in 2015 through the NASA SMEX phase A study, until his death in April 2019. Ted was an inspirational leader and wonderful mentor to many in our team and the solar physics community at large, and will be deeply missed. We are grateful to Jean-Pierre W\"ulser for his major contributions to the design of the MUSE instruments, and to the large team of engineers and scientists that have worked so hard to develop this mission concept. We thank Peter Young who provided the atomic physics data for \ion{Fe}{7} and assisted with estimating MUSE count rates. The MUSE team acknowledges support from NASA contract 80GSFC18C0012 to LMSAL. M.C.M.C., B.D.P., J.M.S. and P.T. acknowledge support by NASA's Heliophysics Grand Challenges Research grant \emph{Physics and Diagnostics of the Drivers of Solar Eruptions} (NNX14AI14G to LMSAL). P.A. acknowledges funding from his STFC Ernest Rutherford Fellowship (grant agreement No. ST/R004285/1). A.D. acknowledges support by NASA’s Heliophysics Technology and Instrument Development for Science grant, \emph{High Efficiency EUV Gratings for Heliophysics}.  We thank Predictive Science Inc for providing the three-dimensional active region simulation data.


%\vspace{5mm}
%\facilities{HST(STIS), Swift(XRT and UVOT), AAVSO, CTIO:1.3m,


%\software{astropy \citep{2013A&A...558A..33A},
%          Cloudy \citep{2013RMxAA..49..137F},
%          SExtractor \citep{1996A&AS..117..393B}
%          }



\appendix
\section{Examples of MUSE spectral profiles}~\label{sec:ap_prof}

\begin{figure*}
    \centering
    \includegraphics[width=0.8\textwidth]{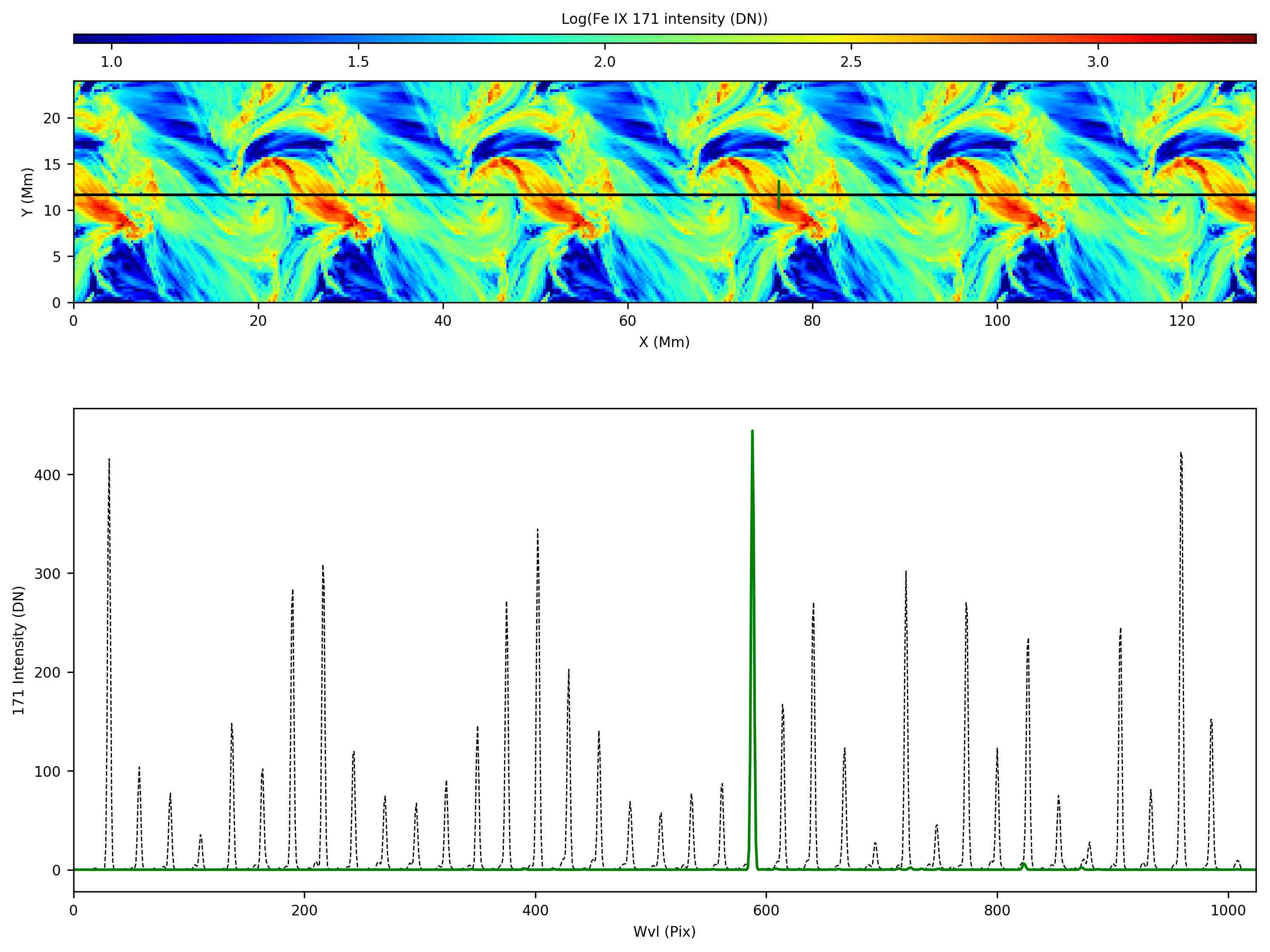}
    \caption{Synthetic MUSE spectra from a horizontal cut from a Bifrost flux emergence simulation. Black dotted lines show the total spectrum, while green full line shows the contribution from one individual slit. Top row shows \ion{Fe}{9} 171\AA\ synthetic MUSE image, bottom shows spectrum in the 171 band. The green line shows \ion{Fe}{9}, cleanly separated from spectra of neighboring slits with very minor contamination (from \ion{Fe}{10} 174.5\AA, see Figure~\ref{fig:fig1}). The numerical domain of this simulation is much smaller than the full MUSE FOV. We tiled the simulation in a periodic fashion multiple times to cover the full FOV. Note that this represents a worst case and unrealistic scenario in terms of potential multi-slit ambiguities, since it implies that MUSE would be observing six bright active region cores occurring at the same time within its FOV. On the real Sun, contamination would thus be reduced compared to this case.}
    \label{fig:fig7}
\end{figure*}

\begin{figure*}
    \centering
    \includegraphics[width=0.8\textwidth]{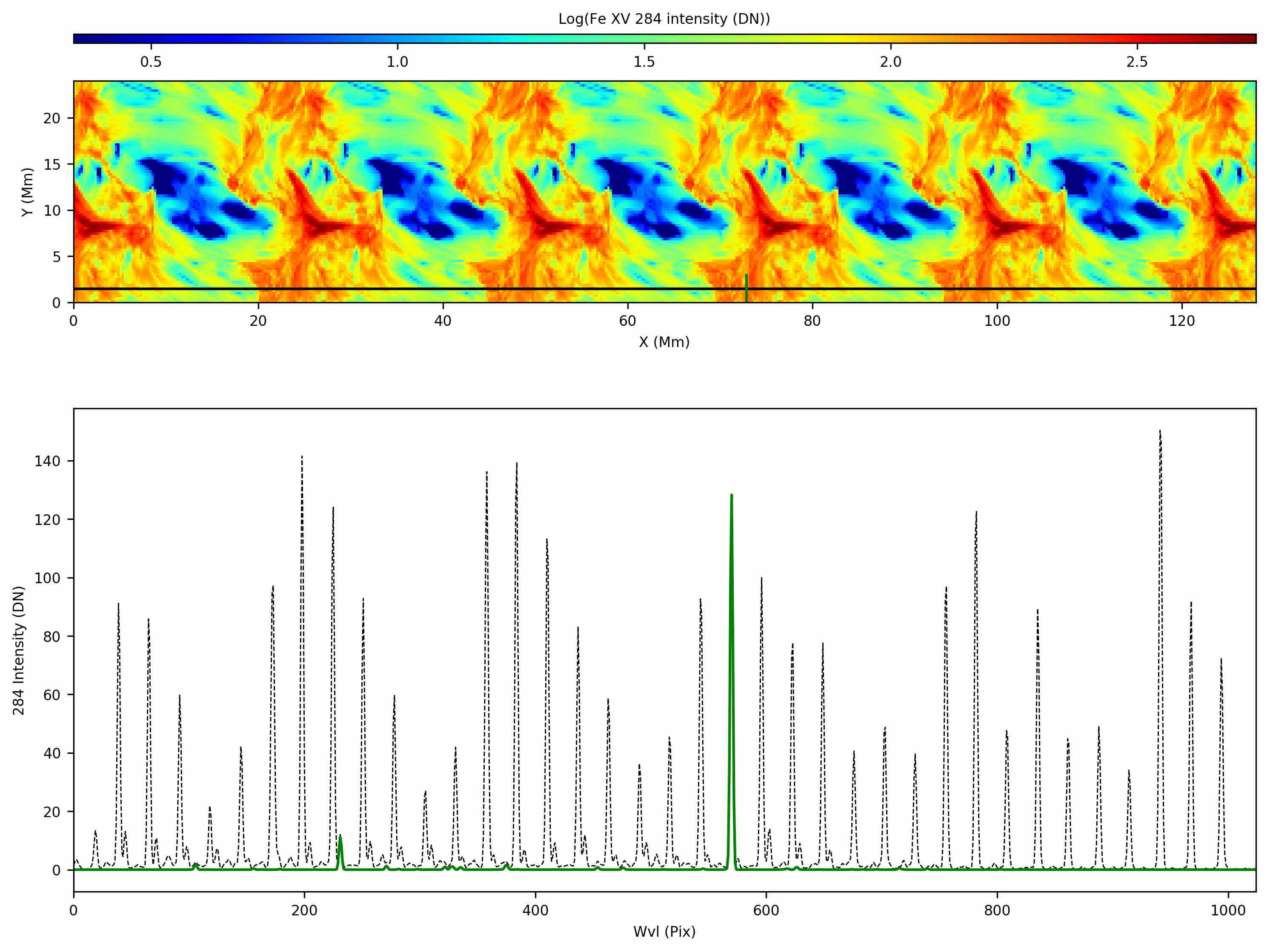}
    \caption{As in figure~\ref{fig:fig7} but for the \ion{Fe}{15} 284\AA.}
    \label{fig:fig8}
\end{figure*}

Several more examples of simulated MUSE spectral line profiles are given in Figures~\ref{fig:fig7} and \ref{fig:fig8}.

\section{Estimates of MUSE count rates} \label{appendix:cnts}
Here we present a brief overview of the methodology we used in order to estimate MUSE count rates for a variety of different solar features.
The calculations were obtained by using several different sources, as explained below. In all cases we transformed line intensities into MUSE count rates by using the effective areas listed in section~\ref{sec:muse_intro}.

\subsection{Active Region}\label{sec:cnts:AR}

In order to estimate the expected MUSE active region count rates we used SDO/AIA and Hinode/EIS data.

\begin{itemize}
    \item{SDO/AIA:} We used a sample of AIA
    observations of 10 different ARs (NOAA 12470, 12473, 12476, 12480, 12487, 12488, 12489, 12490, 12494, 12497; observed between Dec 16 2015 and Feb 9 2016) in non-flaring conditions.  For each AR we used 16 AIA datasets -- each comprising images in the 94\AA, 131\AA, 171\AA, 193\AA, 211\AA, and 335\AA\ —- at a cadence of 2 minutes, therefore covering 30 minutes in total. This temporal sampling allows to take into account the variability of the coronal emission. For each dataset we derived differential emission measure (DEM), pixel by pixel, with the method of \cite{Cheung:AIADEMs}. From the AIA DEMs we derived intensities in the MUSE 108\AA, 171\AA, and 284\AA\ lines, and the 
    195\AA\ coronal imager passband, pixel by pixel, by using CHIANTI emissivities (assuming CHIANTI ionization equilibrium and coronal abundances). We then derived MUSE count rates assuming a pixel dimension of [0.166\arcsec,0.4\arcsec] for the spectrograph, and 
    [0.14\arcsec,0.14\arcsec] for the coronal imager, and MUSE effective areas.  For the 304\AA\ coronal imager passband, given the known inadequacy of the existing atomic database in predicting its emission \citep{Boerner2014}, we have instead used the observed AIA values to directly estimate the MUSE count rates, by scaling them by the relative effective area and pixel size. This seems a reasonable approach given the similarity between the expected 304\AA\ MUSE passband and the SDO/AIA 304\AA\ passband.
    \item{Hinode/EIS:} We measured \ion{Fe}{15} 284\AA, and \ion{Fe}{9} 197.86\AA\ Hinode/EIS emission, to calculate MUSE 171\AA, and 284\AA\ count rates respectively. To estimate count rates in the MUSE 195\AA\ context imager we summed the emission of the strongest lines in the passband, measured by Hinode/EIS (therefore deriving a lower limit to the expected MUSE count rates). The values we obtained for MUSE 171\AA\ and 284\AA\ emission in AR from AIA analysis above are in agreement with an independent estimate we have carried out by using Hinode/EIS spectral observations of ARs (21-Jun-2010 14:24 UT, 10-Aug-2010 22:38UT, 26-Oct-2010 10:49UT, 16-Feb-2011 07:03UT, 15-Apr-2011), and calculating the corresponding predicted MUSE count rates. 
\end{itemize}

\subsection{Quiet Sun, QS Bright Point, Coronal Hole, and CH Bright Point}\label{sec:cnts:QS}

For quiet Sun (QS), coronal hole (CH), and Bright Points (BPs), we derived the MUSE expected counts in the 304\AA\ imager by scaling AIA 304\AA\ counts as described in the previous section for ARs. For 171\AA, 284\AA, and 195\AA\ analogously to the ARs case, we estimated MUSE count rates by measuring \ion{Fe}{15} 284\AA, and \ion{Fe}{9} 197.86\AA\ Hinode/EIS emission, to calculate MUSE 284\AA, and 171\AA\ count rates respectively, while to estimate count rates in the MUSE 195\AA\ coronal imager we summed the emission of the strongest lines in the passband, measured by Hinode/EIS. For these quieter solar features we use the following datasets:

\begin{itemize}
    \item{QS:} 30-Jan-2007 (also analyzed by \citealt{Brooks2009}, and 23-Jan-2008 at 20:50UT
    \item{QS Bright Point:} 23-Jan-2008 at 20:50UT, and 8-Oct-2010 at 10:15UT
    \item{CH and CH Bright Point:} 24-May-2007 at 15:51UT
\end{itemize}
\subsection{M2 flare}\label{sec:cnts:flare}
We derived the MUSE expected counts in the 304\AA\ imager by scaling (as described in sec.~\ref{sec:cnts:AR} for ARs) average AIA 304\AA\ counts for a M2 flare observed by AIA on 2015-09-20 around 18UT.
For 171\AA, 284\AA, and 195\AA\ we estimated MUSE count rates based on the CHIANTI flare DEM (flare.dem; and assuming coronal abundances).
We also estimated MUSE expected count rates using Hinode/EIS observations of the M2 flare on 6-Mar-2012 at 12:38UT.

\section{MUSE Spectral Response Functions} \label{sec:sdc_response}

The MUSE spectral response functions give the detector response of the spectrograph across all 1024 spectral pixels for all three channels to a unit emission measure ($10^{27}$~cm$^{-5}$) of plasma at a specified slit (1 through 37), temperature, and velocity. These are a key and necessary ingredient in order to perform a VDEMS inversion, and more generally, greatly speed up the synthesis of MUSE spectra.  Contribution functions, $G(T)$, for all emission lines in the three passbands are taken from CHIANTI 9.01 (with an updated Fe VII atom provided by Peter Young), specifically: 3329 lines from 100.20 to 120.78\AA, 2302 lines from 167.20 to 187.78\AA, and 2632 lines from 264.40 to 305.55\AA. The reduction in effective area outside of these wavelength ranges is so large that lines beyond these ranges do not contribute significantly. When computing spectra directly from numerical simulations line by line
the inclusion of so many thousands of lines can be computationally burdensome. In contrast, generating a response function cube for a given set of input parameters (plasma pressure, elemental abundances, instrumental parameters) can be performed once to generate spectra more rapidly from many simulations.
To be the most accurate, all the lines are included in the response functions, but it is worth noting that the vast majority of these lines do not contribute significant counts, and it is not necessary to include such a large number of lines for spectral synthesis.
For example, if a threshold in line strength is applied to reduce the number of lines by a factor of 20, the maximum effect on any pixel of any response function is only
or order $10^{-3}$~photon~s$^{-1}$, that is, insignificant.

Instrumental effects, thermal broadening, Doppler shifts, and thermal bremsstrahlung are included in the second step of a two-step process.  The first step is to generate a set of $G(T)$ for a specified plasma pressure and set of abundances as described above.  For the VDEMS inversion studies, the pressures and other parameters used are given in Table 2.
Elemental abundances are generally observed to vary in the solar outer atmosphere according to their first ionization potential (FIP), with low-FIP elements typically enhanced in the corona \citep[e.g.,][]{Meyer1985,Feldman1992,Testa2010SSRv,Testa2015RSPTA}.
In order to span the
maximum range of abundance variations, we used the coronal abundances of \citet{1992PhyS...46..202F} and the photospheric abundances of \citet{2007SSRv..130..105G}. 

In the second step, line intensities are multiplied by the MUSE effective area, thermally broadened gaussian profiles are placed in the appropriate pixels for the given slit and line-of-sight velocity, and then convolved with the instrument spectral response. To facilitate the inclusion of all instrumental effects, this second step is performed with a higher spectral sampling than the MUSE pixels, and then re-binned to the MUSE pixel width.  The instrumental spectral response includes the effects of the optical point spread function (PSF), camera modulation transfer function (MTF), detector charge spreading, and the finite slit width.  Currently, this is implemented with the convolution of a gaussian incorporating the former three effects with a top-hat for the slit width.  Many additional effects can and will be incorporated as needed, for example: measurements and calculations of the PSF and MTF, possibly asymmetric responses, and slit- and wavelength-dependent effective areas and resolutions.  These will be determined through a combination of pre-launch and on-orbit calibrations.

In addition to the instrumental and plasma input parameters discussed above, response functions can be generated with a number of options to facilitate analysis, such as: without bremsstrahlung, without the main lines, with the main lines only, and with a multiplication factor applied to the CHIANTI calculation of \ion{He}{2}~304\AA, as discussed in sections~\ref{sec:purity_sims} and \ref{sec:sdc}.  The MUSE effective area in the 284\AA\ SG passband at 284\AA\ is about 500 times larger than at 304\AA, as the multilayer coating is tuned so that 304\AA\ is suppressed.  Nevertheless, the \ion{He}{2} line can be strong in some locations and appear as a minor contaminant with an offset of 672 pixels or 25.2 slit spacings relative to \ion{Fe}{15}, that is, the faint \ion{He}{2} 304\AA\ signal from slits 1 through 12 will appear near in the 284\AA\ spectra of slits 26 through 37.  The \ion{He}{2} multiplier option allows the user to apply a conservatively large factor to compensate for the fact \ion{He}{2} is typically under-estimated by CHIANTI, or to remove the \ion{He}{2} and add an optically thick calculation separately.  The possible effect of this scaled \ion{He}{2} is included in all spectral purity and SDC analyses in this paper and found not to be a significant effect on the \ion{Fe}{15} line parameters. This is also found to be the case when scaling the \ion{He}{2} intensity to observed signals (see \S~\ref{sec:sdc_target}).

The three main spectral lines as well as the relevant contaminant lines of the selected MUSE spectral windows are weakly pressure dependent. Likewise we find that varying the abundances does not significantly affect the results of the analysis because the dominant contribution in all three bands comes from low-FIP elements that all behave similarly in the coronal context \citep[e.g.,][]{Feldman1992,Testa2010SSRv}. 

A final note on the response functions and the robustness of the SDC method. 
The MUSE passbands are well known and have been observed before, either with EIS (284\AA\ and to some extent 171\AA; \citealt{Brown2008:EIS}), Chandra (108\AA; \citealt{Weisskopf2002:Chandra,Brinkman1987:LETG,Testa2012:Procyon}), and sounding rockets (171\AA; \citealt{2010ApJS..186..222W}). There is the possibility that some weak unknown lines are present in MUSE data. Given the fact that these passbands are well known and the robustness of the SDC to extreme cases like the ones we present in this paper, it is very unlikely that any unknown lines are bright enough to rise to the level of impacting the SDC inversions.  However, even if there were such lines, we note that the MUSE observing con-ops allows on-orbit disambiguation in a statistical sense. Using the Piezo-electric Transducers (PZTs) or by repointing the telescope, we can rapidly place the same solar source under different slits, causing any contaminant lines to be recognized and characterized, even when they are weak or unknown. This is because their position will shift in a predictable fashion depending on which slit produces them. Such lines can even be made to disappear if placed on the leftmost or rightmost slit, depending on whether it is to the blue or red of the main line. By studying the properties of these unknown lines (appearance, Doppler shifts), it is possible to estimate their formation temperature and thus come up with a response function, which should address the issue in this unlikely scenario.



\end{document}